\documentclass[11pt]{article}

\usepackage[dvips]{graphicx}
\usepackage[english]{babel}

\usepackage{amsmath}

\newcommand{\half}{ {\textstyle\frac{1}{2}} }

\newcommand{\re}{\operatorname{Re}}
\newcommand{\im}{\operatorname{Im}}

\newcommand{\gev}{\operatorname{GeV}}

\newcommand{\sign}{\operatorname{sign}}

\newcommand{\ms}{\mskip 1.5mu}

\newcommand{\jpsi}{J\mskip -2mu/\mskip -0.5mu\Psi}

\begin{document}

\begin{flushright}
DESY 07-094
\end{flushright}

\vspace*{3cm}

\begin{center}
\textbf{\LARGE  Dispersion representations for hard exclusive
  processes:\\[0.4em] 
  beyond the Born approximation}

\vspace*{1.6cm}

{\large
M. Diehl$\ms{}^1$
and
D. Yu.\ Ivanov$\ms{}^{2}$
}

\vspace*{0.4cm}

\textsl{%
$^1$ Theory Group, Deutsches Elektronen-Synchroton DESY, 22603
  Hamburg, Germany \\
$^2$ Sobolev Institute of Mathematics, 630090 Novosibirsk, Russia
}

\vspace*{0.8cm}

\textbf{Abstract}\\[10pt]
\parbox[t]{0.9\textwidth}{
  Several hard exclusive scattering processes admit a description in
  terms of generalized parton distributions and perturbative
  hard-scattering kernels.  Both the physical amplitude and the
  hard-scattering kernels fulfill dispersion relations.  We give a
  detailed investigation of their consistency at all orders in
  perturbation theory.  The results shed light on the information
  about generalized parton distributions that can be extracted from
  the real and imaginary parts of exclusive amplitudes.  They also
  provide a practical consistency check for models of these
  distributions in which Lorentz invariance is not exactly satisfied.}
\end{center}

\vspace*{1cm}


\section{Introduction}

Dispersion relations play an important role in the description of
exclusive processes, relating the real and imaginary parts of the
amplitude.  They are for instance required to derive the operator
product expansion for Compton scattering in Bjorken kinematics.  In
this context they have recently been used to establish a
representation of the deeply virtual Compton amplitude which allows
the inclusion of two-loop corrections in a practicable
way~\cite{Kumericki:2007sa}.  In a different context, dispersion
relations have been employed in \cite{Ivanov:2004vd} to simplify the
calculation of the hard-scattering kernels for exclusive quarkonium
production at next-to-leading order.

For hard exclusive processes that can be calculated using collinear
factorization, one may write down dispersion relations both for the
physical process and for the parton-level subprocess.  The question of
consistency between both representations turns out to be nontrivial
and has already been raised in the seminal work \cite{Collins:1996fb}
giving the proof of factorization for meson production.
Important progress has recently been reported in
\cite{Teryaev:2005uj}, where it was shown that this consistency is
ensured by Lorentz invariance in the form of the polynomiality
property for generalized parton distributions (GPDs).  The studies in
\cite{Teryaev:2005uj} were carried out using the Born-level
approximation of the hard-scattering subprocess.  In particular, they
showed that to this accuracy not only the imaginary but also the real
part of the process amplitude can be represented in terms of GPDs
$F(x,\xi,t)$ along the line $x=\xi$ in the $x$--$\xi$ plane.  This
constitutes both a simplification and a limitation for extracting
information on GPDs from hard exclusive amplitudes at leading-order
accuracy.  It is natural to ask how the situation changes when
including radiative corrections to the hard-scattering kernel.

In the present work we therefore investigate dispersion
representations for hard exclusive processes to all orders in
perturbation theory, generalizing the leading-order results derived in
\cite{Teryaev:2005uj}.  In addition we consider in detail the
distributions for polarized quarks and for gluons, for which special
issues arise.  Our paper is organized as follows.  In the next section
we recall a number of results which will be needed in our subsequent
work.  Section~\ref{sec:unpol-q} gives a detailed analysis of
dispersion representations in the unpolarized quark sector.  The
specifics of other distributions are discussed in
Sect.~\ref{sec:other-dist}.  As an application of our results, we
investigate in Sect.~\ref{sec:mc} the model for GPDs proposed by
McDermott, Freund and Strikman \cite{Freund:2002qf}, where
polynomiality is not satisfied.  In Sect.~\ref{sec:sum} we summarize
our findings and draw conclusions.


\section{Some reminders}

Let us begin by recalling some well-known properties of generalized
parton distributions and of dispersion relations, which we will need
in the subsequent sections.


\subsection{Lorentz invariance and crossing properties}

An essential property of generalized parton distributions is the
polynomiality of their Mellin moments.  This property directly follows
from the Lorentz covariance of the operator matrix elements which are
parameterized by GPDs \cite{Ji:1998pc}.  With the conventional
definitions (given e.g.\ in \cite{Diehl:2003ny}) we have for quarks
\begin{align}
  \label{gpd-q-mom}
\int_{-1}^1 dx\, x^{n-1}\, H^q(x,\xi,t) &=
\sum_{\substack{k=0\\{\mathrm{even}}}}^{n-1}
(2\xi)^k\, A^q_{\smash{n,k}}(t) + (2\xi)^{n} C^q_{n}(t)\,,
\nonumber \\
\int_{-1}^1 dx\, x^{n-1}\, E^q(x,\xi,t) &=
\sum_{\substack{k=0\\{\mathrm{even}}}}^{n-1}
(2\xi)^k\, B^q_{\smash{n,k}}(t) - (2\xi)^{n} C^q_{n}(t)\,,
\nonumber \\
\int_{-1}^1 dx\, x^{n-1}\, \widetilde{H}^q(x,\xi,t) &=
\sum_{\substack{k=0\\{\mathrm{even}}}}^{n-1}
(2\xi)^k\, \widetilde{A}^q_{\smash{n,k}}(t)\,,
\nonumber \\
\int_{-1}^1 dx\, x^{n-1}\, \widetilde{E}^q(x,\xi,t) &=
\sum_{\substack{k=0\\{\mathrm{even}}}}^{n-1}
(2\xi)^k\, \widetilde{B}^q_{\smash{n,k}}(t)
\end{align}
with $n\ge 1$, where $C^q_{n}$ is nonzero only for even $n$.  For
gluons we have
\newpage
\begin{align}
  \label{gpd-g-mom}
\int_{0}^1 dx\, x^{n-2}\, H^g(x,\xi,t) &=
\sum_{\substack{k=0\\{\mathrm{even}}}}^{n-2}
(2\xi)^k\, A^g_{\smash{n,k}}(t) + (2\xi)^{n} C^g_{n}(t)\,,
\nonumber \\
\int_{0}^1 dx\, x^{n-2}\, E^g(x,\xi,t) &=
\sum_{\substack{k=0\\{\mathrm{even}}}}^{n-2}
(2\xi)^k\, B^g_{\smash{n,k}}(t) - (2\xi)^{n} C^g_{n}(t)\,,
\nonumber \\
\int_{0}^1 dx\, x^{n-2}\, \widetilde{H}^g(x,\xi,t) &=
\sum_{\substack{k=0\\{\mathrm{even}}}}^{n-1}
(2\xi)^k\, \widetilde{A}^g_{\smash{n,k}}(t)\,,
\nonumber \\
\int_{0}^1 dx\, x^{n-2}\, \widetilde{E}^g(x,\xi,t) &=
\sum_{\substack{k=0\\{\mathrm{even}}}}^{n-1}
(2\xi)^k\, \widetilde{B}^g_{\smash{n,k}}(t)\,,
\end{align}
where $n\ge 2$ is even for $H^g$ and $E^g$ and $n\ge 3$ is odd for
$\widetilde{H}^g$ and $\widetilde{E}^g$.  Since $H^g$, $E^g$ are even
and $\widetilde{H}^g$, $\widetilde{E}^g$ are odd functions of $x$, we
can restrict the integrals in \eqref{gpd-g-mom} to the range $0<x<1$.
The convention for the moment index $n$ is such that quark and gluon
form factors with the same $n$ mix under evolution, i.e.\
$A^q_{\smash{n,k}}$ with $A^g_{\smash{n,k}}$, $B^q_{\smash{n,k}}$ with
$B^g_{\smash{n,k}}$ etc.  The different powers of $x$ in the integrals
\eqref{gpd-q-mom} and \eqref{gpd-g-mom} reflect the different forward
limits of the distributions, e.g.\ $H^q(x,0,0) = q(x)$ and $H^g(x,0,0)
= x g(x)$ for $x>0$.

An important ingredient in the subsequent discussion will be the
high-energy behavior of scattering amplitudes.  According to the
principles of Regge theory, this behavior is connected with the
quantum numbers exchanged in the $t$-channel.
Let us briefly recall how the relevant quantum numbers can be
determined in the context of generalized parton distributions
\cite{Polyakov:1999gs}.  For negative or zero $t$ the form factors
$A^q_{\smash{n,k}}(t)$ etc.\ parameterize the matrix elements of quark
or gluon operators between single-proton states.  Their analytic
continuation to positive $t$ gives the corresponding matrix elements
between the vacuum and a proton-antiproton state.  Decomposing those
matrix elements into contributions with definite angular momentum, one
can associate the form factors with the relevant quantum numbers in
the $t$-channel.  The relevant decomposition for the GPDs of the
proton is given in Chapt.~4.2 of~\cite{Diehl:2003ny}, and we list the
resulting $J^{PC}$ quantum numbers of the $t$-channel exchange in
Table~\ref{tab:JPC}.  {}From this one can readily establish the
exchange quantum numbers for the generalized parton distributions,
which are given in Table~\ref{tab:GPD}.  In particular we see that for
positive charge conjugation there are distributions allowing for
spin-zero exchange.  This corresponds to energy independent
contributions in scattering amplitudes, which play a prominent role in
dispersion relations as we will see.

\begin{table}
  \caption{\label{tab:JPC} Quantum numbers of $t$-channel exchanges for
    the form factors in \protect\eqref{gpd-q-mom} and
    \protect\eqref{gpd-g-mom} as explained in the text.  The entries
    with positive charge conjugation parity $C=+1$ refer to both
    quarks gluons, and those with $C=-1$ only to quarks.}
\renewcommand{\arraystretch}{1.3}
\begin{center}
\begin{tabular}{lcl} \hline\hline
\multicolumn{1}{c}{form factor} &
   $n$ & \multicolumn{1}{c}{$J^{PC}$} \\ \hline
$A_{n,k} + \dfrac{t\rule{0pt}{2ex}}{4m^2}\, B_{n,k}$ &
   even & $0^{++}, 2^{++}, \ldots, (n-k)^{++}$ \\
$C_{n}\rule{0pt}{3ex}$ & even & $0^{++}$ \\
$A_{n,k} + B_{n,k}\rule{0pt}{3.5ex}$ &
   even & $\hspace{2.2em} 2^{++}, \ldots, (n-k)^{++}$ \\
$\widetilde{A}_{n,k}
 + \dfrac{t\rule{0pt}{2ex}}{4m^2}\, \widetilde{B}_{n,k}$ &
   odd  & $0^{-+}, 2^{-+}, \ldots, (n-k-1)^{-+}$ \\
$\widetilde{A}_{n,k}\rule{0pt}{3ex}$ &
   odd  & $1^{++}, 3^{++}, \ldots, (n-k)^{++}$ \\ \hline
$A_{n,k} + \dfrac{t\rule{0pt}{2ex}}{4m^2}\, B_{n,k}$ &
   odd  & $1^{--}, 3^{--}, \ldots, (n-k)^{--}$ \\
$A_{n,k} + B_{n,k}\rule{0pt}{3ex}$ &
   odd  & $1^{--}, 3^{--}, \ldots, (n-k)^{--}$ \\
$\widetilde{A}_{n,k}
  + \dfrac{t\rule{0pt}{2ex}}{4m^2}\, \widetilde{B}_{n,k}$ &
   even & $1^{+-}, 3^{+-}, \ldots, (n-k-1)^{+-}$ \\
$\widetilde{A}_{n,k}\rule{0pt}{3ex}$ &
   even & $\hspace{2.2em} 2^{--}, \ldots, (n-k)^{--}$ \\
\hline\hline
\end{tabular}
\end{center}
\end{table}

\begin{table}
  \caption{\label{tab:GPD} Quantum numbers of $t$-channel exchanges for
    combinations of generalized quark distributions of definite charge
    conjugation parity.  The entries with $C=+1$ also hold for the
    corresponding gluon distributions.}
\renewcommand{\arraystretch}{1.4}
\begin{center}
\begin{tabular}{cl} \hline\hline
distribution & \multicolumn{1}{c}{$J^{PC}$} \\ \hline
$H^q(x,\xi,t) - H^q(-x,\xi,t)$ & $0^{++}, 2^{++}, \ldots$ \\
$E^q(x,\xi,t) - E^q(-x,\xi,t)$ & $0^{++}, 2^{++}, \ldots$ \\
$\widetilde{H}^q(x,\xi,t) + \widetilde{H}^q(-x,\xi,t)$ &
  $1^{++}, 3^{++}, \ldots$ \\
$\widetilde{E}^q(x,\xi,t) + \widetilde{E}^q(-x,\xi,t)$ &
  $0^{-+}, 1^{++}, 2^{-+}, 3^{++}, \ldots$ \\ \hline
$H^q(x,\xi,t) + H^q(-x,\xi,t)$ & $1^{--}, 3^{--}, \ldots$ \\
$E^q(x,\xi,t) + E^q(-x,\xi,t)$ & $1^{--}, 3^{--}, \ldots$ \\
$\widetilde{H}^q(x,\xi,t) - \widetilde{H}^q(-x,\xi,t)$ &
  $2^{--}, 4^{--}, \ldots$ \\
$\widetilde{E}^q(x,\xi,t) - \widetilde{E}^q(-x,\xi,t)$ &
  $1^{+-}, 2^{--}, 3^{+-}, 4^{--}, \ldots$ \\
\hline\hline
\end{tabular}
\end{center}
\end{table}

A way to ensure polynomiality of the moments \eqref{gpd-q-mom} is the
double distribution representation
\cite{Musatov:1999xp,Polyakov:1999gs}
\begin{align}
  \label{dd-q1}
H^q(x,\xi,t) &= H^q_{f}(x,\xi,t)
  + \sign(\xi)\, D^q\Bigl(\frac{x}{\xi},t\Bigr) \,,
&
E^q(x,\xi,t) &= E^q_{k}(x,\xi,t)
  - \sign(\xi)\, D^q\Bigl(\frac{x}{\xi},t\Bigr)
\end{align}
with
\begin{align}
  \label{dd-q2}
H^q_{f}(x,\xi,t) &= \int d\beta\, d\alpha\;
  \delta(x-\alpha\xi-\beta)\, f^q(\beta,\alpha,t) \,,
\nonumber \\
E^q_{k}(x,\xi,t) &= \int d\beta\, d\alpha\;
  \delta(x-\alpha\xi-\beta)\, k^q(\beta,\alpha,t) \,,
\end{align}
where $f^q$ and $k^q$ are commonly referred to as double distributions
and $D^q$ as the $D$-term.  The support region of
$f^q(\beta,\alpha,t)$ and $k^q(\beta,\alpha,t)$ is the rhombus
$|\alpha| + |\beta| \le 1$, whereas $D^q(\alpha,t)$ has support for
$|\alpha| < 1$ and is odd in $\alpha$.  More general representations
have been discussed in the literature
\cite{Polyakov:1999gs,Teryaev:2001qm,Tiburzi:2004qr} but will not be
needed in the following.  For gluons one has
\begin{align}
  \label{dd-g1}
H^g(x,\xi,t) &= H^g_{f}(x,\xi,t)
  + |\xi|\, D^g\Bigl(\frac{x}{\xi},t\Bigr) \,,
&
E^g(x,\xi,t) &= E^g_{k}(x,\xi,t)
 - |\xi|\, D^g\Bigl(\frac{x}{\xi},t\Bigr)
\end{align}
with
\begin{align}
  \label{dd-g2}
H^g_{f}(x,\xi,t) &= \int d\beta\, d\alpha\;
  \delta(x-\alpha\xi-\beta)\, \beta f^g(\beta,\alpha,t) \,,
\nonumber \\
E^g_{k}(x,\xi,t) &= \int d\beta\, d\alpha\;
  \delta(x-\alpha\xi-\beta)\, \beta k^g(\beta,\alpha,t) \,.
\end{align}
The support properties of $f^g$, $k^g$ and $D^g$ are as for their
quark counterparts, and $D^g(\alpha,t)$ is even in~$\alpha$.  One
readily finds that the Mellin moments of the $D$-term are related to
the form factors $C_n(t)$ as
\begin{align}
  \label{D-C}
\int_{-1}^1 d\alpha\, \alpha^{n-1} D^q(\alpha,t) &= 2^n C^q_n(t) \,,
&
\int_{0}^1 d\alpha\,  \alpha^{n-2} D^g(\alpha,t) &= 2^n C^g_n(t) \,.
\end{align}
The polarized quark distributions $\widetilde{H}^q$ and
$\widetilde{E}^q$ have double distribution representations analogous
to \eqref{dd-q1} and \eqref{dd-q2} but without a $D$-term, since the
highest power appearing in their Mellin moments \eqref{gpd-q-mom} is
$\xi^{n-1}$ instead of $\xi^n$.  We will discuss the case of
$\widetilde{H}^g$ and $\widetilde{E}^g$ in Section~\ref{sec:pol-g}.


\subsection{Dispersion relations}

The exclusive processes we consider in this work are deeply virtual
Compton scattering (DVCS) and light meson production,
\begin{align}
  \label{procs}
\gamma^*(q) + p(p) &\to \gamma(q') + p(p') \,,
&
\gamma^*(q) + p(p) &\to M(q') + p(p') \,,
\end{align}
where four-momenta are indicated in parentheses.  Our arguments can be
extended to the production of heavy mesons like the $\jpsi$, but we
shall not dwell on this here.  Since the processes in \eqref{procs}
involve particles with nonzero spin, the appropriate quantities for
discussing dispersion relations are invariant amplitudes, which have
simple analyticity and crossing properties.  An explicit decomposition
for Compton scattering can be found in \cite{Belitsky:2001ns}, where
these invariant amplitudes are called Compton form factors.

To describe the kinematics of \eqref{procs} we use the Mandelstam
variables $s=(p+q)^2$, $t=(p-p')^2$, $u=(p-q')^2$.  Consider now an
invariant amplitude $\mathcal{F}^{[\sigma]}(\nu,t)$ with definite
signature $\sigma$ under $s \leftrightarrow u$ crossing, so that
\begin{equation}
  \label{crossing}
\mathcal{F}^{[\sigma]}(-\nu,t)
  = \sigma \mathcal{F}^{[\sigma]}(\nu,t) \,,
\end{equation}
where $2\nu = s-u$.  We will work in kinematics where $t \le 0$ and
external photons are on shell or have spacelike virtuality, so that
the imaginary part of the amplitude is due to the $s$-channel
discontinuity for $\nu>0$ and to the $u$-channel discontinuity for
$\nu<0$. The fixed-$t$ dispersion relation with no subtraction then
reads
\begin{equation}
  \label{disp-unsub}
\re\mathcal{F}^{[\sigma]}(\nu,t)
= \frac{1}{\pi} \int_{\nu_{th}}^{\infty} d\nu'\,
  \im\mathcal{F}^{[\sigma]}(\nu',t)
  \left[ \frac{1}{\nu'-\nu} + \sigma \frac{1}{\nu'+\nu} \right] \,,
\end{equation}
where $\nu_{th}$ is the value of $\nu$ at threshold.  Here and in the
following Cauchy's principal value prescription is understood for the
singularities at $\nu'= \pm\nu$ of the dispersion integral.  For the
dispersion relation \eqref{disp-unsub} to be valid, the integral of
$\mathcal{F}^{[\sigma]}(\nu',t)$ times the term in square brackets
must vanish when taken over an infinite semicircle in the $\nu'$
plane.  This requires
\begin{align}
  \label{no-sub}
\mathcal{F}^{[+]}(\nu,t)
  & \underset{|\nu| \to\infty}{\to} 0 \,, & 
\nu^{-1} \mathcal{F}^{[-]}(\nu,t)
  & \underset{|\nu| \to\infty}{\to} 0 \,.
\end{align}
A dispersion relation with one subtraction,
\begin{align}
  \label{disp-sub}
& \re\mathcal{F}^{[\sigma]}(\nu,t) - \re\mathcal{F}^{[\sigma]}(\nu_0,t)
\nonumber \\
& \qquad
= \frac{1}{\pi} \int_{\nu_{th}}^{\infty} d\nu'\,
  \im\mathcal{F}^{[\sigma]}(\nu',t)
  \left[ \frac{1}{\nu'-\nu} + \sigma \frac{1}{\nu'+\nu} 
       - \frac{1}{\nu'-\nu_0} - \sigma \frac{1}{\nu'+\nu_0}\right] ,
\end{align}
is valid if
\begin{align}
  \label{one-sub}
\nu^{-2} \mathcal{F}^{[+]}(\nu,t) 
   & \underset{|\nu| \to\infty}{\to} 0 \,,
\end{align}
whereas for $\sigma=-1$ we have the same condition \eqref{no-sub} as
with no subtraction.

We will study dispersion relations for the processes \eqref{procs} in
the Bjorken limit of large $-q^2$ at fixed $q^2/\nu$ and $t$.  It is
then useful to trade $\nu$ for the scaling variable
\begin{equation}
  \label{xi-def}
\xi = -\frac{(q+q')^2}{2\ms (p+p')\cdot (q+q')}
    = -\frac{q^2}{s-u} = -\frac{q^2}{2\nu} \,,
\end{equation}
where we have neglected $q'^2$ and $t$ compared with $q^2$ in the
numerator.  The factorization theorems established in
\cite{Collins:1996fb,Collins:1998be} state that in the Bjorken limit
certain invariant amplitudes become dominant and can be written as the
convolution of partonic hard-scattering kernels with generalized quark
or gluon distributions (and the light-cone distribution amplitude of
the produced meson).\footnote{%
  Up to terms suppressed by inverse powers of $\sqrt{-q^2}$, the
  leading invariant amplitudes for DVCS correspond to transverse
  photon polarization and those for meson production to longitudinal
  photon and meson polarization in the collision c.m.}
To establish dispersion relations we will need information on the
high-energy behavior of these amplitudes.  Empirically the small-$x$
behavior of the usual quark and gluon distributions, obtained from
fits mainly to inclusive deep inelastic scattering data, is well
described by a power law.  With currently used models for generalized
parton distributions, based either on double distributions or on
Gegenbauer moments, one finds a corresponding power-law behavior for
the invariant amplitudes of DVCS
\cite{Belitsky:2001ns,Freund:2001rk,Mueller:2005ed,Kumericki:2007sa}
and of meson production \cite{Goloskokov:2006hr}.  Whether this
correspondence may be model-independent is not known, see the
discussion in Sect.~3.2 of \cite{Mueller:2005ed}.  We will take it as
a guideline in the following, bearing in mind that deviations between
the power laws of parton densities and exclusive amplitudes (or
deviations from a strict power behavior in the asymptotic limit) do
not invalidate our dispersion relations as long as the invariant
amplitudes do not grow faster than the critical power of energy
specified in \eqref{no-sub} and~\eqref{one-sub}.


\section{Unpolarized quark distributions}
\label{sec:unpol-q}

In this section we discuss in detail the contribution of unpolarized
quark distributions to the leading invariant amplitudes for DVCS or
meson production.  Here and in the following we decompose all
amplitudes into terms of definite signature $\sigma$.  According to
the factorization theorem we can write
\begin{equation}
  \label{fact-orig}
\mathcal{F}^{q [\sigma]}(\xi,t,q^2)
= \int_{-1}^1 dx\, \frac{1}{\xi}\,
  C^{q [\sigma]}\Bigl( \frac{x}{\xi}, q^2 \Bigr)\, F^q(x,\xi,t)
\end{equation}
with $F^q = H^q, E^q$.  For simplicity we have omitted the dependence
on the renormalization and factorization scales; in the following will
also omit the arguments $q^2$ in the hard-scattering kernel\footnote{%
  We refer to $C^{q [\sigma]}$ as hard-scattering kernel for ease of
  language, keeping in mind that for meson production it is more
  precisely the convolution of a hard-scattering kernel with the meson
  distribution amplitude.}
and $t$ in the generalized parton distributions.  The hard-scattering
kernel satisfies the symmetry relation
\begin{equation}
  \label{kernel-symm}
  C^{q [\sigma]}\Bigl( -\frac{x}{\xi} \Bigr)
= -\sigma\ms C^{q [\sigma]}\Bigl( \frac{x}{\xi} \Bigr) \,,
\end{equation}
so that the factorization formula can be written as
\begin{equation}
  \label{fact-form}
\mathcal{F}^{q [\sigma]}(\xi)
= \int_{0}^1 dx\, \frac{1}{\xi}\,
  C^{q [\sigma]}\Bigl( \frac{x}{\xi} \Bigr)\,
  F^{q [\sigma]}(x,\xi)
\end{equation}
in terms of the combinations
\begin{equation}
  \label{gpd-sign-def}
F^{q [\sigma]}(x,\xi) = F^q(x,\xi) - \sigma F^q(-x,\xi)
\end{equation}
for quark exchange of definite signature.  We remark that $F^{q[+]}$
corresponds to positive and $F^{q[-]}$ to negative charge conjugation
parity in the $t$-channel.  With the relation
\begin{equation}
  \label{gpd-sign-prop}
F^{q [\sigma]}(x,-\xi) = F^{q [\sigma]}(x,\xi)
\end{equation}
from time reversal invariance one finds $\mathcal{F}^{q
[\sigma]}(-\xi) = \sigma \mathcal{F}^{q [\sigma]}(\xi)$ as required.
In the Bjorken limit the Mandelstam variables for the hard-scattering
subprocess are given by
\begin{align}
\hat{s} &= x s + \half (1-x)\ms q^2 \,,
&
\hat{u} &= x u + \half (1-x)\ms q^2 \,,
\end{align}
so that one has
\begin{equation}
\frac{x}{\xi} = - \frac{\hat{s} - \hat{u}}{q^2} \,.
\end{equation}
To leading order (LO) in $\alpha_s$ the kernel reads
\begin{align}
  \label{born-kernel}
  C^{q [\sigma]}(\omega) &\,\propto\,
  \frac{1}{1-\omega-i\epsilon} - \sigma \frac{1}{1+\omega-i\epsilon} \,,
&
  \im C^{q [\sigma]}(\omega) &\,\propto\, \pi\ms
  \bigl[\ms \delta(\omega-1) - \sigma \delta(\omega+1) \ms\bigr]
\end{align}
for both DVCS and meson production, where we have omitted any global
factors which are irrelevant for our discussion of fixed-$t$
dispersion relations here.
At higher orders in $\alpha_s$ one finds branch cuts in the $\hat{s}$
and $\hat{u}$ channels for $\omega>1$ and $\omega<-1$, respectively.
For the dispersion relations to be discussed shortly, we need to know
the behavior of the kernels for $|\omega| \to \infty$.  The NLO
kernels for DVCS can be found in \cite{Belitsky:2005qn}, and those for
meson production in \cite{Ivanov:2004zv}.  For negative signature, one
finds $C^{q [-]}(\omega) \sim \omega^{-1}$ up to logarithms for both
DVCS and meson production.  For positive signature, the NLO
corrections give $C^{q [+]}(\omega) \sim \omega^{-1}$ for DVCS, and
$C^{q[+]}(\omega) \sim \omega^0$ for meson production, again up to
logarithms.  The power behavior as $\omega^0$ is due to two-gluon
exchange in the $t$-channel.  For DVCS such graphs only start at NNLO,
so that at this level one will also have $C^{q[+]}(\omega) \sim
\omega^0$.  This change in energy behavior between NLO and NNLO is the
same as in the hard-scattering kernels for inclusive deep inelastic
scattering \cite{Zijlstra:1992qd}, obtained from $\gamma^* p\to
\gamma^* p$ in forward kinematics via the optical theorem.  In fact,
the kernels for DVCS and for deep inelastic scattering are intimately
related, see e.g.~\cite{Kumericki:2007sa,Belitsky:2005qn}.


\subsection{Dispersion relations}
\label{sec:disp-rels}

The invariant amplitude satisfies a fixed-$t$ dispersion relation.
Using $1/\xi = -2\nu /q^2$ as energy variable and making one
subtraction, one has
\begin{equation}
  \label{disp-rel-1}
   \re \mathcal{F}^{q [\sigma]}(\xi)
 - \re \mathcal{F}^{q [\sigma]}(\xi_0)
= \frac{1}{\pi} \int_1^\infty d\omega'\,
    \im \mathcal{F}^{q [\sigma]}(1/\omega')
    \left[ \frac{1}{\omega' - 1/\xi} + \sigma \frac{1}{\omega' + 1/\xi}
           - \{ \xi\to \xi_0 \} \right] \,,
\end{equation}
where $\xi_0$ denotes the subtraction point and the Cauchy principal
value prescription is understood at $\omega' = \pm 1/\xi$.  As is
appropriate in the Bjorken limit, we have neglected $t$ and the hadron
masses when determining the lower limit of the $\omega'$ integration.

According to the discussion at the end of the previous section, the
validity of a dispersion relation with one subtraction requires that
$\xi^2 \mathcal{F}^{q[+]}(\xi)$ and $\xi \mathcal{F}^{q[-]}(\xi)$
vanish for $\xi\to 0$, whereas an unsubtracted dispersion relation
would require $\mathcal{F}^{q[+]}(\xi) \to 0$ in the same limit.
Given the phenomenological observed small-$x$ behavior of valence and
sea quark distributions, we expect a small-$\xi$ behavior
$\mathcal{H}^{q [\sigma]}(\xi) \sim \xi^{-\alpha}$ with $1< \alpha <2$
for $\sigma=+1$ and $0< \alpha <1$ for $\sigma=-1$.  For $\sigma=+1$
we hence do require one subtraction in the dispersion relation.  We
have also taken one subtraction for $\sigma=-1$ although this would
not be necessary.  We shall see that our final results for negative
signature would be the same with no subtraction.  According to
Table~\ref{tab:GPD} the distributions $H^{q [\sigma]}$ and $E^{q
  [\sigma]}$ involve the same quantum numbers in the $t$-channel, and
we therefore expect that the high-energy behavior of $\mathcal{H}^{q
  [\sigma]}$ and $\mathcal{E}^{q [\sigma]}$ is similar.

Inserting the factorization formula \eqref{fact-form} into
\eqref{disp-rel-1} and using that $C^{q [\sigma]}(\omega)$ has a
vanishing imaginary part for $|\omega| <1$, one obtains
\begin{align}
  \label{disp-rel-1.5}
& \re \mathcal{F}^{q [\sigma]}(\xi)
 - \re \mathcal{F}^{q [\sigma]}(\xi_0)
\nonumber \\
&\quad
 = \frac{1}{\pi} \int_1^\infty d\omega'  \int_{1/\omega'}^1 dx\,
    \omega' \im C^{q [\sigma]}(x \omega')\, F^{q [\sigma]}(x,1/\omega')
    \left[ \frac{1}{\omega' - 1/\xi} + \sigma \frac{1}{\omega' + 1/\xi}
           - \{ \xi\to \xi_0 \} \right]
\nonumber \\
&\quad
= \frac{1}{\pi} \int_1^\infty d\omega  \int_0^1 dx\,
    \frac{\omega}{x^2}\, \im C^{q [\sigma]}(\omega)\,
    F^{q [\sigma]}\Bigl(x,\frac{x}{\omega}\Bigr)\,
    \left[ \frac{1}{\omega/x - 1/\xi} + \sigma \frac{1}{\omega/x + 1/\xi}
           - \{ \xi\to \xi_0 \} \right] \,,
\end{align}
where from the second to the third line we have changed the order of
integration, $\smash{\int_1^\infty d\omega' \int_{1/\omega'}^1 dx} =
\int_0^1 dx \int_{1/x}^\infty d\omega'$, substituted $\omega = x
\omega'$, and changed the order of integration again.  Straightforward
algebra finally gives
\begin{align}
  \label{disp-rel-2}
& \re \mathcal{F}^{q [\sigma]}(\xi)
  = \re \mathcal{F}^{q [\sigma]}(\xi_0)
\nonumber \\
&\quad
+ \frac{1}{\pi} \int_1^\infty d\omega\, \im C^{q [\sigma]}(\omega)
   \int_0^1 dx\, F^{q [\sigma]}\Bigl(x,\frac{x}{\omega}\Bigr)\,
    \left[ \frac{1}{\omega\xi - x} - \sigma \frac{1}{\omega\xi + x}
         - \frac{1}{\omega\xi_0 - x}
                  + \sigma \frac{1}{\omega\xi_0 + x} \right] \,.
\end{align}
Note that $\im C^{q [\sigma]}(\omega)$ contains terms proportional to
$\delta(\omega - 1)$, as is already seen in the leading-order
expression \eqref{born-kernel}.  These terms are understood to be
included in the integration over $\omega$ in \eqref{disp-rel-2}.
A remark is in order on the behavior of the integrand for $x\to 0$.
Let us first consider the case $F^{q [\sigma]} = H^{q [\sigma]}$.  It
is natural to expect that $H^{q [\sigma]}(x,x/\omega)$ has a singular
behavior for $x\to 0$ that is similar to the forward distribution
$q(x) + \sigma \bar{q}(x)$.  With the small-$x$ behavior of quark
densities obtained in typical phenomenological analyses, one then has
an integrable singularity of $H^{q [-]}(x,x/\omega)$, whereas the
corresponding singularity of $H^{q [+]}(x,x/\omega)$ is stronger than
$x^{-1}$ but weaker than $x^{-2}$.  For $\sigma=+1$ the expression in
square brackets in \eqref{disp-rel-2} is however proportional to $x$,
so that the integrand is again sufficiently well behaved at $x=0$.  A
similar discussion can be given for $E^{q [\sigma]}(x,x/\omega)$,
assuming that its small-$x$ behavior is similar to the one of $H^{q
[\sigma]}(x,x/\omega)$.

We now discuss the dispersion relation for the hard-scattering kernel
itself.  Notice that according to \eqref{kernel-symm} the kernel $C^{q
  [\sigma]}$ has opposite symmetry behavior under crossing than the
corresponding process amplitude $\mathcal{F}^{q [\sigma]}$, so that
$C^{q[+]}$ satisfies a negative-signature dispersion relation and
$C^{q[-]}$ a positive-signature one.  With the large-$\omega$ behavior
discussed after \eqref{born-kernel} we hence need no subtraction in
either case and can write
\begin{equation}
\re C^{q [\sigma]}\Bigl( \frac{x}{\xi} \Bigr)
= \frac{1}{\pi} \int_1^\infty d\omega\,
   \im C^{q [\sigma]}(\omega)
   \left[ \frac{1}{\omega - x/\xi} - \sigma \frac{1}{\omega + x/\xi}
   \right] \,,
\end{equation}
where again the Cauchy principal value prescription is implied at
$\omega = \pm x/\xi$.  Insertion into the factorization formula
\eqref{fact-form} yields
\begin{equation}
  \label{disp-rel-3}
\re \mathcal{F}^{q [\sigma]}(\xi)
= \frac{1}{\pi} \int_1^\infty d\omega\, \im C^{q [\sigma]}(\omega)
  \int_0^1 dx\, F^{q [\sigma]}(x,\xi)
   \left[ \frac{1}{\omega\xi - x} - \sigma \frac{1}{\omega\xi + x}
   \right] \,.
\end{equation}
This can in particular be used to evaluate the subtraction constant
$\re \mathcal{F}^{q [\sigma]}(\xi_0)$ in \eqref{disp-rel-2}, which
then reads
\begin{align}
  \label{disp-rel-4}
\re \mathcal{F}^{q [\sigma]}(\xi)
 = \frac{1}{\pi} \int_1^\infty d\omega\, \im C^{q [\sigma]}(\omega)
   & \int_0^1 dx\, \biggl\{
    F^{q [\sigma]}\Bigl(x,\frac{x}{\omega}\Bigr)\,
    \left[ \frac{1}{\omega\xi - x} - \sigma \frac{1}{\omega\xi + x}
    \right]
\nonumber \\[0.2em]
&\, + \biggl[ F^{q [\sigma]}(x,\xi_0)
         - F^{q [\sigma]}\Bigl(x,\frac{x}{\omega}\Bigr) \biggr]
    \left[ \frac{1}{\omega\xi_0 - x} - \sigma \frac{1}{\omega\xi_0 + x}
    \right] \biggr\} \,.
\end{align}
Notice that the terms in the second line give the amplitude in the
limit $\xi\to \infty$, which corresponds to the point $s = u =
q^2/2$ in the unphysical region.  The negative-signature amplitude
must vanish at this point for symmetry reasons.  Comparison of the
$\xi_0$ independent terms in \eqref{disp-rel-1.5} and
\eqref{disp-rel-2} shows that an unsubtracted dispersion relation for
$\mathcal{F}^{q[-]}(\xi)$ has indeed the same form as
\eqref{disp-rel-4} without the terms in the second line.  The same is
however not true for $\mathcal{F}^{q[+]}(\xi)$.

Consistency of the representations \eqref{disp-rel-3} and
\eqref{disp-rel-4} implies
\begin{multline}
  \label{disp-master}
\frac{1}{\pi}
 \int_1^\infty d\omega\, \im C^{q [\sigma]}(\omega) \int_{-1}^1 dx\,
  \biggl[ F^{q}(x,\xi)
        - F^{q}\Bigl(x,\frac{x}{\omega}\Bigr) \biggr]
  \left[ \frac{1}{\omega\xi - x} - \sigma \frac{1}{\omega\xi + x}
  \right]
\\
= \frac{1}{\pi}
 \int_1^\infty d\omega\, \im C^{q [\sigma]}(\omega) \int_{-1}^1 dx\,
  \biggl[ F^{q}(x,\xi_0)
        - F^{q}\Bigl(x,\frac{x}{\omega}\Bigr) \biggr]
  \left[ \frac{1}{\omega\xi_0 - x} - \sigma \frac{1}{\omega\xi_0 + x}
  \right] \,,
\end{multline}
i.e.\ the l.h.s.\ must be independent of $\xi$.  In
\eqref{disp-master} we have restored the integration over negative $x$
and traded $F^{q [\sigma]}$ for $F^q$, making use of the symmetry
relation \eqref{gpd-sign-prop}.  The Cauchy principal value
prescription should be applied at $x=0$ if $\sigma=-1$, so that a
possible nonintegrable singularity of the $\sigma=+1$ part of
$F^q(x,x/\omega) = \half \bigl[ F^{q [+]}(x,x/\omega) + F^{q
[-]}(x,x/\omega) \bigl] $ cancels under the integral because it is
antisymmetric in~$x$.
At this point we can make two comments:
\begin{enumerate}
\item To leading order in $\alpha_s$ the dispersion representation
  \eqref{disp-rel-2} involves only distributions $F^{q
  [\sigma]}(x,\xi)$ at the point $x=\xi$ because of the simple form
  \eqref{born-kernel} of the hard-scattering kernel, as was found in
  \cite{Teryaev:2005uj}.  At higher orders in $\alpha_s$ it involves
  however the distributions in the full DGLAP region $|x| \ge \xi$.
  Knowledge of $F^{q [\sigma]}(x,x)$ for all $x$ is hence only
  sufficient to reconstruct the amplitude (up to a subtraction term)
  at leading order in the strong coupling.  The reconstruction is
  however possible to any order in $\alpha_s$ without direct knowledge
  of the distributions in the ERBL region $|x| < \xi$.
\item The consistency of dispersion relations for the process
  amplitude and for the hard-scattering kernel was already discussed
  in the context of the factorization proof in \cite{Collins:1996fb}.
  Translated into our notation, the analog of our
  eq.~\eqref{disp-rel-2} in that work was mistakenly written with
  $F^{q [\sigma]}(x,\xi)$ instead of $F^{q [\sigma]}(x,x/\omega)$ and
  without a subtraction term, so that consistency with
  \eqref{disp-rel-3} was trivial.  The correct consistency relation
  \eqref{disp-master} follows from the polynomiality property of GPDs,
  as we now show.
\end{enumerate}


\subsection{Consequences for generalized parton distributions}
\label{sec:gpd-conseq}

Clearly \eqref{disp-master} is satisfied if
\begin{equation}
  \label{I-def}
\mathcal{I}^{q [\sigma]}(\omega)
= \int_{-1}^1 dx\,
  \biggl[ F^{q}(x,\xi)
        - F^{q}\Bigl(x,\frac{x}{\omega}\Bigr) \biggr]
  \left[ \frac{1}{\omega\xi - x} - \sigma \frac{1}{\omega\xi + x}
  \right]
\end{equation}
is independent of $\xi$ for all $\omega \ge 1$.  To show that this is
the case, we Taylor expand $F^q(x,x/\omega)$ in its second argument,
\begin{align}
\mathcal{I}^{q [\sigma]}(\omega)
&= \frac{1}{\omega} \sum_{n=1}^\infty \frac{1}{n!}\,
   \Bigl( \frac{\partial}{\partial\eta} \Bigr)^n
   \int_{-1}^1 dx\,
     \Bigl(\frac{x}{\omega}-\xi\Bigr)^{n-1}
     F^q(x,\eta)\, \Bigr|_{\eta=\xi}
\nonumber \\
&\quad + \sigma
   \frac{1}{\omega} \sum_{n=1}^\infty \frac{1}{n!}\,
   \Bigl( \frac{\partial}{\partial\eta} \Bigr)^n
   \int_{-1}^1 dx\,
     \Bigl(\frac{x}{\omega}+\xi\Bigr)^{n-1}
     F^q(x,\eta)\, \Bigr|_{\eta=-\xi} \,,
\end{align}
where we have interchanged the order of differentiation and
integration.  For definiteness we consider first the case $F^q = H^q$.
Using the polynomiality property \eqref{gpd-q-mom} and the fact that
$C^q_n$ is only nonzero for even $n$, we find
\begin{align}
  \label{H-result-1}
\mathcal{I}^{q [+]}(\omega)
&= 2 \sum_{\substack{n=2\\{\mathrm{even}}}}^\infty
    \left(\frac{2}{\omega}\right)^{n} C^q_{n}\, ,
&
\mathcal{I}^{q [-]}(\omega) &= 0 \,,
\end{align}
which is independent of $\xi$ as required.  We recall that we have
suppressed the dependence on $t$ in the distributions $F^q$, as well
as in the form factors $C^q_n$.  Alternatively one may use the double
distribution representation in \eqref{dd-q1} and \eqref{dd-q2}.  One
readily finds that the double distribution part of $\mathcal{I}^{q
[\sigma]}$ is zero, with
\begin{align}
  \label{int-H-dd}
& \int_{-1}^1 dx\,
  \biggl[ H^{q}_{f}(x,\xi)
        - H^{q}_{f}\Bigl(x,\frac{x}{\omega}\Bigr) \biggr]\,
  \frac{1}{\omega\xi - x}
\nonumber \\
&\quad
= \int_{-1}^1 dx \int d\beta\, d\alpha\, f^q(\beta,\alpha)\,
  \biggl[ \delta(x - \alpha\xi - \beta)
        - \delta\bigl(x\ms [1- \tfrac{\alpha}{\omega}] - \beta\bigr)
  \biggr]\,
  \frac{1}{\omega\xi - x}
\nonumber \\
&\quad
= \int d\beta\, d\alpha\, f^q(\beta,\alpha)\, \left[
  \frac{1}{\omega\xi - \alpha\xi - \beta}
  - \frac{1/ (1-\frac{\alpha}{\omega})}{\omega\xi
    - \beta/ (1-\frac{\alpha}{\omega})} \right]
\nonumber \\
&\quad = 0  \phantom{\biggl[ \biggr]}
\end{align}
and an analogous relation for the term with $1/(\omega\xi + x)$.  The
only nonzero contribution to $\mathcal{I}^{q [\sigma]}$ comes hence
from the $D$-term
\begin{align}
  \label{H-result-2}
\mathcal{I}^{q [+]}(\omega)
 &= \sign(\xi) \int_{-1}^1 dx\, D^q\Bigl(\frac{x}{\xi}\Bigr)\,
  \left[ \frac{1}{\omega\xi - x} - \frac{1}{\omega\xi + x}
  \right]
= 2 \int_{-1}^1 d\alpha\, \frac{D^q(\alpha)}{\omega-\alpha} \,,
\nonumber \\[0.2em]
\mathcal{I}^{q [-]}(\omega) &= 0 \,,
\end{align}
where we have used the support and symmetry properties of
$D^q(\alpha)$ stated after \eqref{dd-q2}.  Expanding
$1/(\omega-\alpha)$ in a geometric series and using \eqref{D-C} one
readily sees that \eqref{H-result-1} and \eqref{H-result-2} are
equivalent.  For the case $F^q = E^q$ the discussion proceeds in full
analogy, with the opposite sign of $C^q_n$ in \eqref{H-result-1} and
of $D^q$ in \eqref{H-result-2}.
As a corollary one finds the integral relations
\begin{align}
  \label{int-relations-1}
& \int_{-1}^1 dx\,
  \biggl[ H^{q}(x,\xi)
        - H^{q}\Bigl(x,\frac{x}{\omega}\Bigr) \biggr]
  \left[ \frac{1}{\omega\xi - x} - \frac{1}{\omega\xi + x}
  \right]
\nonumber \\
&\quad = - \int_{-1}^1 dx\,
  \biggl[ E^{q}(x,\xi)
        - E^{q}\Bigl(x,\frac{x}{\omega}\Bigr) \biggr]
  \left[ \frac{1}{\omega\xi - x} - \frac{1}{\omega\xi + x}
  \right]
= 2 \int_{-1}^1 dx\, \frac{D^q(x)}{\omega-x}
\intertext{and}
  \label{int-relations-2}
& \int_{-1}^1 dx\,
  \biggl[ H^{q}(x,\xi) 
       - H^{q}\Bigl(x,\frac{x}{\omega}\Bigr) \biggr]
  \left[ \frac{1}{\omega\xi - x} + \frac{1}{\omega\xi + x}
  \right]
\nonumber \\
&\qquad = \int_{-1}^1 dx\,
  \biggl[ E^{q}(x,\xi)
        - E^{q}\Bigl(x,\frac{x}{\omega}\Bigr) \biggr]
  \left[ \frac{1}{\omega\xi - x} + \frac{1}{\omega\xi + x}
  \right]
= 0 \,.
\end{align}
They reflect the polynomiality properties of the distributions and in
this sense are non-trivial consequences of Lorentz invariance.  Using
them to evaluate the $\xi_0$ dependent terms in \eqref{disp-rel-4}
gives
\begin{align}
  \label{disp-rel-5a}
\re\mathcal{H}^{q [+]}(\xi) &=
  \frac{1}{\pi} \int_1^\infty d\omega\, \im C^{q [+]}(\omega)
  \int_{-1}^1 dx\, \biggl\{
    H^{q}\Bigl(x,\frac{x}{\omega}\Bigr)\,
    \left[ \frac{1}{\omega\xi - x} - \frac{1}{\omega\xi + x}
    \right] + \frac{2 D^q(x)}{\omega-x} \biggr\} \,,
\nonumber \\
\re\mathcal{E}^{q [+]}(\xi) &=
  \frac{1}{\pi} \int_1^\infty d\omega\, \im C^{q [+]}(\omega)
  \int_{-1}^1 dx\, \biggl\{
    E^{q}\Bigl(x,\frac{x}{\omega}\Bigr)\,
    \left[ \frac{1}{\omega\xi - x} - \frac{1}{\omega\xi + x}
    \right] - \frac{2 D^q(x)}{\omega-x} \biggr\} \,,
\\
\intertext{and}
  \label{disp-rel-5b}
\re\mathcal{H}^{q [-]}(\xi) &=
  \frac{1}{\pi} \int_1^\infty d\omega\, \im C^{q [-]}(\omega)
  \int_{-1}^1 dx\,
    H^{q}\Bigl(x,\frac{x}{\omega}\Bigr)\,
    \left[ \frac{1}{\omega\xi - x} + \frac{1}{\omega\xi + x}
    \right]
\end{align}
with an analogous representation for $\mathcal{E}^{q [-]}(\xi)$.  We
note that according to our comment after \eqref{disp-rel-4} one has
\begin{equation}
  \label{amp-inf}
\lim_{\xi\to\infty} \mathcal{F}^{q [\sigma]}(\xi)
= \frac{1}{\pi} \int_1^\infty d\omega\, \im C^{q [\sigma]}(\omega)\,
  \mathcal{I}^{q [\sigma]}(\omega) \,.
\end{equation}
For $\sigma=-1$ this is zero, and in fact we could have immediately
obtained \eqref{disp-rel-5b} from an unsubtracted dispersion relation,
where the $\xi_0$ dependent terms in \eqref{disp-rel-4} are absent as
remarked earlier.  For $\sigma=+1$, the subtraction term in the
dispersion relation \eqref{disp-rel-2} is fixed by the $D$-term if one
takes $\xi_0\to \infty$.  In the leading-order approximation for the
hard-scattering kernel this was already observed in
\cite{Teryaev:2005uj}, and for the general case in
\cite{Kumericki:2007sa}.  According to Table~\ref{tab:JPC} the
$D$-term parameterizes a part of $H^q$ and $E^q$ which is associated
with spin-zero exchange in the $t$-channel.\footnote{%
  Note that this is not restricted to the exchange of spin-zero
  resonances.  In the context of chiral dynamics
  \protect\cite{Ando:2006sk} the dominant exchange is in fact given by
  two pions in an $S$-wave.}
{}From \eqref{fact-form} one readily finds that its contribution to
the invariant amplitudes $\mathcal{H}^{q [+]}(\xi)$ and
$\mathcal{E}^{q [+]}(\xi)$ is energy-independent and purely real.


\subsection{The Compton amplitude with both photons off shell}
\label{sec:ddvcs}

So far we have discussed deeply virtual Compton scattering, $\gamma^*
p\to \gamma p$, where the photon in the final state is on shell, and
obtained the integral relations \eqref{int-relations-1} and
\eqref{int-relations-2} for the generalized parton distributions.  It
is natural to ask whether any further relations can be derived by
considering dispersion relations for the Compton amplitude
\begin{equation}
  \label{ddvcs}
\gamma^*(q) + p(p) \to \gamma^*(q') + p(p')
\end{equation}
with both photons off shell.  For $q^2<0$ and $q'^2>0$ this process
can be studied experimentally, with the timelike final-state photon
decaying into a lepton pair \cite{Guidal:2002kt}.  The analyticity
properties of the amplitude are however more complicated in this case,
because there are simultaneous branch cuts in $s$ and $q'^2$ or in $u$
and $q'^2$.  Instead we consider the case where both $q^2$ and $q'^2$
are spacelike, so that the only singularities are in $s$ and $u$, as
in the previous subsections.  We have two scaling variables
\begin{align}
\xi &= -\frac{(q+q')^2}{2\ms (p+p')\cdot (q+q')}
     = -\frac{q^2 + q'^2}{s-u} \,,
&
\vartheta &= \frac{q^2 - q'^2}{q^2 + q'^2} \,,
\end{align}
where in the second expression for $\xi$ we have neglected $t$
compared with $q^2 + q'^2$.  For $\vartheta=1$ we recover the case of
DVCS, whereas with two spacelike photon virtualities we have $-1 <
\vartheta < 1$.  In the Bjorken limit of large $-q^2$ at fixed $\xi$,
$\vartheta$ and $t$ one has a factorization formula for the invariant
amplitudes
\begin{equation}
  \label{fact-ddvcs}
\mathcal{F}^{q [\sigma]}(\xi,\vartheta,t,q^2)
= \int_{-1}^1 dx\, \frac{1}{\xi}\,
  C^{q [\sigma]}\Bigl( \frac{x}{\xi}, \vartheta, q^2 \Bigr)\,
  F^q(x,\vartheta\ms \xi,t)
\end{equation}
with $F^q = H^q, E^q$ as before.  We will again omit the arguments
$q^2$ and $t$ in the following.  The Mandelstam variables of the hard
subprocess now read
\begin{align}
\hat{s} &= x s + \half (1-x)\ms (q^2 + q'^2) \,,
&
\hat{u} &= x u + \half (1-x)\ms (q^2 + q'^2) \,,
\end{align}
in the Bjorken limit, so that $x/\xi = -(\hat{s} - \hat{u}) /(q^2 +
q'^2)$.  For a dispersion relation at fixed $t$ and fixed photon
virtualities, $\vartheta$ plays the role of a constant parameter, and
we can use $1/\xi$ and $x/\xi$ as respective energy variable of the
overall process and the hard subprocess.  In the Bjorken limit the
corresponding amplitudes have branch cuts in $1/\xi$ or $x/\xi$ from
$1$ to $\infty$ and from $-\infty$ to $-1$.  The hard-scattering
kernel has the symmetry
\begin{equation}
  C^{q [\sigma]}\Bigl( -\frac{x}{\xi}, \vartheta \Bigr)
= -\sigma\ms C^{q [\sigma]}\Bigl( \frac{x}{\xi} , \vartheta \Bigr)
\end{equation}
in analogy to \eqref{kernel-symm}.  At leading order in $\alpha_s$ it
reads
\begin{align}
  C^{q [\sigma]}(\omega,\vartheta) &\,\propto\,
  \frac{1}{1-\omega-i\epsilon} - \sigma \frac{1}{1+\omega-i\epsilon} \,,
&
  \im C^{q [\sigma]}(\omega,\vartheta) &\,\propto\, \pi\ms
  \bigl[\ms \delta(\omega-1) - \sigma \delta(\omega+1) \ms\bigr] \,,
\end{align}
and at higher orders it has the same high-$\omega$ behavior as
discussed for DVCS after \eqref{born-kernel}.  In other words, the
high-energy behavior of the hard-scattering kernel for the virtual
Compton amplitude \eqref{ddvcs} remains unchanged if $q'^2\to 0$.
Similarly, the small-$\xi$ behavior of $\mathcal{F}^{q
  [\sigma]}(\xi,\vartheta)$ is as discussed for DVCS after
\eqref{disp-rel-1}.  One can thus derive dispersion relations for the
invariant amplitude and for the hard-scattering kernel as in
Sect.~\ref{sec:disp-rels} and finds
\begin{align}
  \label{ddvcs-rel-3}
\re \mathcal{F}^{q [\sigma]}(\xi,\vartheta)
&= \frac{1}{\pi} \int_1^\infty d\omega\,
  \im C^{q [\sigma]}(\omega,\vartheta)
  \int_0^1 dx\, F^{q [\sigma]}(x,\vartheta\ms \xi)
   \left[ \frac{1}{\omega\xi - x} - \sigma \frac{1}{\omega\xi + x}
   \right]
\intertext{and}
  \label{ddvcs-rel-4}
\re \mathcal{F}^{q [\sigma]}(\xi,\vartheta)
&= \frac{1}{\pi} \int_1^\infty d\omega\,
   \im C^{q [\sigma]}(\omega,\vartheta)
   \int_0^1 dx\, \biggl\{
    F^{q [\sigma]}\Bigl(x,\vartheta\ms \frac{x}{\omega}\Bigr)\,
    \left[ \frac{1}{\omega\xi - x} - \sigma \frac{1}{\omega\xi + x}
    \right]
\nonumber \\[0.3em]
& \hspace{6em}
+ \biggl[ F^{q [\sigma]}(x,\vartheta\ms \xi_0)
        - F^{q [\sigma]}\Bigl(x,\vartheta\ms \frac{x}{\omega}\Bigr)
      \biggr]
    \left[ \frac{1}{\omega\xi_0 - x} - \sigma \frac{1}{\omega\xi_0 + x}
    \right] \biggr\} \,.
\end{align}
These relations read exactly as their counterparts \eqref{disp-rel-3}
and \eqref{disp-rel-4} for DVCS, except that the second argument of
$F^{q [\sigma]}$ is now multiplied with $\vartheta$ and that $C^{q
[\sigma]}$ depends on $\vartheta$ as well.  The consistency of
\eqref{ddvcs-rel-3} and \eqref{ddvcs-rel-4} is ensured if
\begin{equation}
\int_{-1}^1 dx\,
  \biggl[ F^q(x,\vartheta\ms \xi)
        - F^q\Bigl(x,\vartheta\ms \frac{x}{\omega}\Bigr) \biggr]
  \left[ \frac{1}{\omega\xi - x} - \sigma \frac{1}{\omega\xi + x} \right]
\end{equation}
is independent of $\xi$ for all $\omega\ge 1$.  Rescaling $\xi' =
\vartheta\ms \xi$ and $\omega' = \omega/\vartheta$, we readily see
that this in ensured by the $\xi$-independence of the integral
$\mathcal{I}^{q [\sigma]}(\omega)$ in \eqref{I-def}, which we have
already established.  Thus the dispersion relations for doubly virtual
Compton scattering give no new relations for GPDs.  Of course, one
obtains dispersion representations for $\mathcal{H}^{q
[\sigma]}(\xi,\vartheta)$ and $\mathcal{E}^{q
[\sigma]}(\xi,\vartheta)$ as in \eqref{disp-rel-5a} and
\eqref{disp-rel-5b}, with $\vartheta$ as an additional argument in
$C^{q [\sigma]}$ and with the replacements $H^q(x, x/\omega) \to
H^q(x, \vartheta\ms x/\omega)$, $E^q(x, x/\omega) \to E^q(x,
\vartheta\ms x/\omega)$ and $D^q(x)\, (\omega - x)^{-1} \to D^q(x)\,
(\omega/\vartheta - x)^{-1}$.

Let us now consider the case $q = q'$, relevant for deep inelastic
scattering, where we have $\xi = x_B$ and $\vartheta=0$.  The
representations \eqref{ddvcs-rel-3} and \eqref{ddvcs-rel-4} are then
trivially consistent, because the second argument of $F^{q [\sigma]}$
is zero everywhere.  In other words, the usual parton densities
appearing in inclusive processes do not depend on an external
kinematical variable, unlike the generalized parton distributions
appearing in exclusive processes.

In the following section we investigate the contributions from
polarized quark distributions and from gluons to DVCS and to meson
production.  The results we will obtain can readily be generalized to
the Compton amplitude with two spacelike photons.  We note that for
the unpolarized quark distributions we have just considered, only the
amplitudes with $\sigma=+1$ appear in Compton scattering, whereas
$\sigma=-1$ is relevant for the polarized quark and gluon
distributions.


\section{Polarized and gluon distributions}
\label{sec:other-dist}

Contributions from polarized quarks and from unpolarized or polarized
gluons to invariant amplitudes can be treated in a similar manner as
the case of unpolarized quarks in the previous section.
Particularities arise for each of the distributions, which we will now
discuss in turn.


\subsection{Polarized quark distributions}
\label{sec:pol-q}

Let us first investigate invariant amplitudes involving polarized
quark distributions, which appear in both DVCS and in the production
of pseudoscalar mesons.  The factorization formula reads as in
\eqref{fact-orig}, where now $F^q = \widetilde{H}^q$ or
$\widetilde{E}^q$.  We define combinations $\widetilde{H}^{q
  [\sigma]}$ and $\widetilde{E}^{q [\sigma]}$ of definite signature as
in \eqref{gpd-sign-def}, and the relations \eqref{kernel-symm} to
\eqref{gpd-sign-prop} are again valid.  Note that, in contrast to
their unpolarized counterparts, $\widetilde{H}^{q[+]}$ and
$\widetilde{E}^{q[+]}$ correspond to negative charge conjugation and
$\widetilde{H}^{q[-]}$ and $\widetilde{E}^{q[-]}$ to positive charge
conjugation in the $t$-channel.  The leading-order expression of the
hard-scattering kernel for DVCS and for meson production is the same
as in \eqref{born-kernel}.  At NLO one finds a large-$\omega$ behavior
$C^{q [\sigma]}(\omega) \sim \omega^{-1}$ up to logarithms in both
cases.  Note that $t$-channel two-gluon exchange in the polarized
sector does not give rise to a power behavior as $\omega^0$.  This is
also explicitly seen in the NNLO kernels for inclusive deep inelastic
scattering \cite{Zijlstra:1993sh}.

For the polarized quark and antiquark densities we assume that $x
\Delta q(x)$ and $x \Delta\bar{q}(x)$ vanish at $x\to 0$, as it is
found in global fits and required for the existence of the moments
$\smash{\int_0^1} dx\, \Delta q(x)$ and $\int_0^1 dx\,
\Delta\bar{q}(x)$.  One should then have a small-$\xi$ behavior $\xi
\smash{\widetilde{\mathcal{H}}^{q [\sigma]}} \to 0$ for both positive
and negative signature, so that the once-subtracted dispersion
relation \eqref{disp-rel-1} is valid.  The argument proceeds as in
Sections~\ref{sec:disp-rels} and \ref{sec:gpd-conseq}.  According to
\eqref{gpd-q-mom} the $x^{n-1}$ moment of $\widetilde{H}^q(x,\xi)$ has
$\xi^{n-1}$ as highest power, so that the integral
$\mathcal{I}^{[\sigma]}(\omega)$ in \eqref{I-def} is zero for both
$\sigma=+1$ and $\sigma=-1$ in this case.  We therefore obtain the
integral relation
\begin{equation}
  \label{rel-tilde}
\int_{-1}^1 dx\, \biggl[ \widetilde{H}^{q}(x,\xi)
        - \widetilde{H}^{q}\Bigl(x,\frac{x}{\omega}\Bigr) \biggr]
  \left[ \frac{1}{\omega\xi - x} \pm \frac{1}{\omega\xi + x}
  \right] = 0
\end{equation}
and dispersion representations
\begin{align}
  \label{disp-rel-tilde}
\re\widetilde{\mathcal{H}}^{q [\sigma]}(\xi) &=
  \frac{1}{\pi} \int_1^\infty d\omega\, \im C^{q [\sigma]}(\omega)
  \int_{-1}^1 dx\,
    \widetilde{H}^{q}(x,\xi)\,
    \left[ \frac{1}{\omega\xi - x} -\sigma \frac{1}{\omega\xi + x}
    \right]
\nonumber \\
&=  \frac{1}{\pi} \int_1^\infty d\omega\, \im C^{q [\sigma]}(\omega)
  \int_{-1}^1 dx\,
    \widetilde{H}^{q}\Bigl(x,\frac{x}{\omega}\Bigr)\,
    \left[ \frac{1}{\omega\xi - x} -\sigma \frac{1}{\omega\xi + x}
    \right] \,.
\end{align}
We further find that $\widetilde{\mathcal{H}}^{q [\pm]}(\xi) \to 0$
for $\xi\to\infty$.  As in the unpolarized case, we could have
obtained the second representation in \eqref{disp-rel-tilde} from a
dispersion relation without subtraction in the case $\sigma=-1$.
For $\sigma=+1$, the high-energy behavior of the invariant amplitude
does however require one subtraction, even though the subtraction term
is zero when taking the subtraction point $\xi_0 \to \infty$.  An
unsubtracted dispersion relation for positive signature would differ
from \eqref{disp-rel-tilde}, as remarked after~\eqref{disp-rel-4}.

For $\widetilde{E}^{q [\sigma]}$ the situation is more involved.
According to Table~\ref{tab:GPD} this distribution admits more
$t$-channel exchanges than $\widetilde{H}^{q [\sigma]}$, so that the
small-$\xi$ behavior of $\widetilde{\mathcal{E}}^{q [\sigma]}(\xi)$
and $\widetilde{\mathcal{H}}^{q [\sigma]}(\xi)$ may be different.  In
particular there is a known spin-zero exchange contribution to
$\widetilde{E}^{q [-]}$, which is due to pion exchange and dominates
the distributions for $u$ and $d$ quarks at small $t$
\cite{Mankiewicz:1998kg,Goeke:2001tz}.  It reads
\begin{equation}
  \label{pion-pole}
\widetilde{E}^{u}_\pi(x,\xi,t) = - \widetilde{E}^{d}_\pi(x,\xi,t)
  = \frac{c}{m_\pi^2-t}\, \frac{1}{|\xi|}\,
    \phi_\pi\Bigl( \frac{x}{\xi} \Bigr) \,,
\end{equation}
where the constant $c$ can be calculated in chiral perturbation theory
\cite{Ando:2006sk} and the light-cone distribution amplitude
$\phi_\pi(\alpha)$ of the pion is an even function with support for
$|\alpha| < 1$.  Inserting this into the factorization formula
\eqref{fact-orig} one obtains a contribution going like $\xi^{-1}$ to
the invariant amplitudes $\widetilde{\mathcal{E}}^{u[-]}$ and
$\widetilde{\mathcal{E}}^{d[-]}$.  This rises too strongly at $\xi\to
0$ for the once-subtracted dispersion relations we have used so far.
At this point we notice that due to the prefactor in its definition,
the distribution $\widetilde{E}^q$ always contributes to matrix
elements as $\xi \widetilde{E}^q$, and correspondingly it is $\xi
\widetilde{\mathcal{E}}^{q [\sigma]}$ which appears in physical
scattering amplitudes.  Note that because of its prefactor $\xi
\widetilde{\mathcal{E}}^{q[-]}(\xi)$ is even in $\xi$ and thus has
positive instead of negative signature.  The pion exchange term
\eqref{pion-pole} gives a $\xi$ independent contribution to $\xi
\widetilde{\mathcal{E}}^{q[-]}(\xi)$, as it should be for spin-zero
exchange.  We can thus write down a once-subtracted dispersion
relation for $\xi \widetilde{\mathcal{E}}^{q [\sigma]}(\xi)$, assuming
only that its small-$\xi$ behavior is less singular than $\xi^{-2}$
for $\sigma=-1$ and less singular than $\xi^{-1}$ for $\sigma=+1$.
The analog of \eqref{disp-rel-3} is now
\begin{equation}
  \label{et-disp-rel-3}
\re \xi \widetilde{\mathcal{E}}^{q [\sigma]}(\xi)
= \frac{1}{\pi} \int_1^\infty d\omega\, \im C^{q [\sigma]}(\omega)
  \int_{-1}^1 dx\; \xi \widetilde{E}^{q}(x,\xi)
   \left[ \frac{1}{\omega\xi - x}
          - \sigma \frac{1}{\omega\xi + x}
   \right] \,,
\end{equation}
and the analog of \eqref{disp-rel-4} reads
\begin{align}
  \label{et-disp-rel-3.5}
\re \xi \widetilde{\mathcal{E}}^{q [\sigma]}(\xi)
 = \frac{1}{\pi} \int_1^\infty &
   d\omega\, \im C^{q [\sigma]}(\omega)
   \int_{-1}^1 dx\, \biggl\{
   \xi_0\ms \widetilde{E}^{q}(x,\xi_0)
   \left[ \frac{1}{\omega\xi_0 - x}
            - \sigma \frac{1}{\omega\xi_0 + x} \right] 
\nonumber \\[0.2em]
& + \frac{x}{\omega}\ms
    \widetilde{E}^{q}\Bigl(x,\frac{x}{\omega}\Bigr)\,
    \left[ \frac{1}{\omega\xi - x}
           + \sigma \frac{1}{\omega\xi + x}
           - \frac{1}{\omega\xi_0 - x}
           - \sigma \frac{1}{\omega\xi_0 + x} \right] \biggr\} \,,
\end{align}
which can be rewritten as
\begin{align}
  \label{et-disp-rel-4}
\re \xi \widetilde{\mathcal{E}}^{q [\sigma]}(\xi)
 = \frac{1}{\pi} \int_1^\infty d\omega\, \im C^{q [\sigma]}(\omega) &
   \int_{-1}^1 dx\, \biggl\{ \ms
    \xi \widetilde{E}^{q}\Bigl(x,\frac{x}{\omega}\Bigr)\,
    \left[ \frac{1}{\omega\xi - x}
           - \sigma \frac{1}{\omega\xi + x}
    \right]
\nonumber \\[0.2em]
& + \xi_0 \biggl[ \widetilde{E}^{q}(x,\xi_0)
    - \widetilde{E}^{q}\Bigl(x,\frac{x}{\omega}\Bigr) \biggr]
    \left[ \frac{1}{\omega\xi_0 - x}
            - \sigma \frac{1}{\omega\xi_0 + x}
    \right] \biggr\} \,.
\end{align}
With the methods of Sect.~\ref{sec:gpd-conseq} one finds
\begin{equation}
  \label{et-int-relations}
\int_{-1}^1 dx\, \biggl[ \widetilde{E}^{q}(x,\xi)
        - \widetilde{E}^{q}\Bigl(x,\frac{x}{\omega}\Bigr) \biggr]
  \left[ \frac{1}{\omega\xi - x} \pm \frac{1}{\omega\xi + x}
  \right] = 0 \,,
\end{equation}
which ensures consistency of the two dispersion representations and
allows us to omit the second line of \eqref{et-disp-rel-4}.  We thus
find that the analog of the representations \eqref{disp-rel-tilde}
also holds for $\widetilde{\mathcal{E}}^{q[\sigma]}$.  In the case
$\sigma=+1$, where spin-zero exchange does not contribute, we could
indeed have obtained this result from a once-subtracted dispersion
relation for $\widetilde{\mathcal{E}}^{q[\sigma]}$.

Notice that the terms in the second line of \eqref{et-disp-rel-4} need
not give $\xi \widetilde{\mathcal{E}}^{q [\sigma]}(\xi)$ at the
unphysical point $\xi\to\infty$, in contrast to the case discussed
after \eqref{disp-rel-4}.  In fact $\xi \widetilde{\mathcal{E}}^{q
[-]}(\xi)$ is nonzero at this point. Taylor expanding $1/(\omega\xi -
x)$ and $1 /(\omega\xi + x)$ in \eqref{et-disp-rel-3} and using the
polynomiality relation~\eqref{gpd-q-mom} one readily finds
\begin{align}
  \label{Etilde-asy-q}
\lim_{\xi\to\infty} \xi \widetilde{\mathcal{E}}^{q [-]}(\xi)
  &= \frac{1}{\pi} \int_1^\infty d\omega\, \im C^{q [-]}(\omega)\,
     \sum_{\substack{n=1\\{\mathrm{odd}}}}^\infty
      \left(\frac{2}{\omega}\right)^{n} \widetilde{B}^q_{n,n-1} \,,
&
\lim_{\xi\to\infty} \xi \widetilde{\mathcal{E}}^{q [+]}(\xi)
  &= 0 \,.
\end{align}
In Table~\ref{tab:JPC} we see that the form factors
$\widetilde{B}^q_{n,n-1}(t)$ are associated with pure spin-zero
exchange.  At small $t$ they are dominated by the the pion-exchange
term \eqref{pion-pole}.  Having support only in the ERBL region $|x| <
\xi$, this term does not contribute to the imaginary part of
$\xi\widetilde{\mathcal{E}}^{q[-]}(\xi)$, and one may wonder how it
can appear in the representation \eqref{et-disp-rel-4} for the real
part.  The answer is that it induces a contribution proportional to
$\delta(x)$ in $\widetilde{E}^{q}(x, x/\omega)$.  To see this we
observe that the double distribution generating \eqref{pion-pole} has
the form $\delta(\beta)\, e_\pi(\alpha,t)$, where we have abbreviated
$e_\pi(\alpha,t) = c\ms (m_\pi^2 -t)^{-1} \phi_\pi(\alpha)$.  For
$\omega \ge 1$ one then has
\begin{align}
  \widetilde{E}^u_{\pi}\Bigl( x,\frac{x}{\omega},t \Bigr)
 &= \int d\beta\, d\alpha\; 
    \delta\bigl(x\ms [1- \tfrac{\alpha}{\omega}] - \beta\bigr)\, 
    \delta(\beta)\, e_\pi(\alpha,t)
 = \delta(x)\, \omega \int_{-1}^1 d\alpha\,
   \frac{e_\pi(\alpha,t)}{\omega-\alpha} \,.
\end{align}
One may avoid this $\delta(x)$ contribution by taking the limit
$\xi_0\to \infty$ in \eqref{et-disp-rel-3.5}, which yields
\begin{align}
  \label{et-alt}
\re \xi \widetilde{\mathcal{E}}^{q [\sigma]}(\xi)
&= \lim_{\xi_0 \to\infty} 
   \xi_0\ms \widetilde{\mathcal{E}}^{q [\sigma]}(\xi_0)
\nonumber \\
&\quad
 + \frac{1}{\pi} \int_1^\infty d\omega\, \im C^{q [\sigma]}(\omega)
   \int_{-1}^1 dx\;
     \frac{x}{\omega}\ms \widetilde{E}^{q}\Bigl(x,\frac{x}{\omega}\Bigr)\,
     \left[ \frac{1}{\omega\xi - x}
            + \sigma \frac{1}{\omega\xi + x} \right]
\end{align}
with the subtraction term given in \eqref{Etilde-asy-q}.  The
$\delta(x)$ contribution in $\widetilde{E}^{q}(x, x/\omega)$ is now
removed by the extra factor $x$ and instead appears explicitly in the
subtraction term.


\subsection{Unpolarized gluon distributions}
\label{sec:unpol-g}

The contribution from unpolarized gluon distributions to invariant
amplitudes can be written as
\begin{equation}
  \label{fact-g}
\mathcal{F}^{g}(\xi)
= \int_{-1}^1 dx\, \frac{1}{\xi}\,
  C^{g}\Bigl( \frac{x}{\xi} \Bigr)\,
  \frac{F^{g}(x,\xi)}{x} \,,
\end{equation}
where $F^g(x,\xi) = H^g(x,\xi), E^g(x,\xi)$ is even in $x$ and in
$\xi$, and the hard-scattering kernel $C^g(\omega)$ is odd in
$\omega$.  The singularity introduced by the factor $1/x$ is spurious
because $C^{g}(\omega) \sim \omega$ at $\omega\to 0$.  For vector
meson production, the hard-scattering kernel reads
\begin{align}
  \label{born-kernel-g}
  C^{g}(\omega) &\,\propto\,
  \frac{1}{1-\omega-i\epsilon} - \frac{1}{1+\omega-i\epsilon} \,,
&
  \im C^{g}(\omega) &\,\propto\, \pi\ms
  \bigl[\ms \delta(\omega-1) - \delta(\omega+1) \ms\bigr]
\end{align}
at LO in $\alpha_s$, whereas for DVCS the kernel for gluon
distributions starts only at NLO.  The high-$\omega$ behavior of
$C^g(\omega)$ at higher orders is the same as discussed for
$C^{q[+]}(\omega)$ after \eqref{born-kernel}.  We assume a small-$x$
behavior like $g(x) \sim x^{-\alpha}$ with $\alpha< 2$ for the
unpolarized gluon density.  The small-$\xi$ behavior
$\mathcal{H}^g(\xi) \sim \xi^{-\alpha}$ is then less singular than
$\xi^{-2}$ and hence admits a once-subtracted dispersion relation.

The symmetry properties of $C^g(\omega)$ and of $x^{-1}\,
F^{g}(x,\xi)$ are identical to those of $C^{q[+]}(\omega)$ and
$F^{q[+]}(x,\xi)$ in the unpolarized quark sector, so that the
dispersion relations for the process amplitude and for the
hard-scattering kernel read exactly as for unpolarized quark
distributions in \eqref{disp-rel-3} and \eqref{disp-rel-4} if one
replaces $\mathcal{F}^{q[+]} \to \mathcal{F}^g$, $C^{q[+]} \to C^g$
and $F^{q[+]} \to 2 x^{-1} F^g$.  Consistency of these dispersion
relations is ensured if
\begin{equation}
  \label{I-g-def}
\mathcal{I}^{g}(\omega)
= \int_{-1}^1 \frac{dx}{x}\,
  \biggl[ F^{g}(x,\xi)
        - F^{g}\Bigl(x,\frac{x}{\omega}\Bigr) \biggr]
  \left[ \frac{1}{\omega\xi - x} - \frac{1}{\omega\xi + x}
  \right]
\end{equation}
is independent of $\xi$.  Using the symmetry properties of $F^g$ we
can replace $1 /(\omega\xi - x) - 1 /(\omega\xi + x)$ by $2
/(\omega\xi - x)$ under the integral, with the principal value
prescription taken to regularize the singularity at $x=0$.  Repeating
the procedure of Section~\ref{sec:gpd-conseq} we Taylor expand
$F^g(x,x/\omega)$ in its second argument and obtain
\begin{equation}
\mathcal{I}^{g}(\omega)
= \frac{2}{\omega} \sum_{n=1}^\infty \frac{1}{n!}\,
   \Bigl( \frac{\partial}{\partial\eta} \Bigr)^n
   \int_{-1}^1 \frac{dx}{x}\,
     \Bigl(\frac{x}{\omega}-\xi\Bigr)^{n-1}
     F^g(x,\eta)\, \Bigr|_{\eta=\xi} \,.
\end{equation}
Since $F^g(x,\eta)$ is even in $x$, a nonzero integral is only
obtained from the odd powers of $x$ in the expansion of $(x/\omega -
\xi)^{n-1}$, so that the factor $x^{-1}$ in the integrand is
canceled.  Using the polynomial property \eqref{gpd-g-mom} one
finally obtains
\begin{equation}
  \label{I-g-C}
\mathcal{I}^g(\omega)
= 4 \sum_{\substack{n=2\\{\mathrm{even}}}}^\infty
    \left(\frac{2}{\omega}\right)^{n} C^g_{n}
\end{equation}
for $F^g = H^g$, which is independent of $\xi$ as required.
Alternatively, one may insert \eqref{dd-g1} and \eqref{dd-g2} into
\eqref{I-g-def}.  For the double distribution part of $H^g$ this gives
\begin{align}
  \label{I-g-dd}
& \int_{-1}^1 \frac{dx}{x}\,
  \biggl[ H^{g}_{f}(x,\xi)
        - H^{g}_{f}\Bigl(x,\frac{x}{\omega}\Bigr) \biggr]\,
  \biggl[ \frac{1}{\omega\xi - x} - \frac{1}{\omega\xi + x}
  \biggr]
\nonumber \\
&\quad
= 2 \int_{-1}^1 dx \int d\beta\, d\alpha\; \beta f^g(\beta,\alpha)\,
  \biggl[ \delta(x - \alpha\xi - \beta)
        - \delta\bigl(x\ms [1- \tfrac{\alpha}{\omega}] - \beta\bigr)
  \biggr]\,
  \frac{1}{x\ms (\omega\xi - x)}
\nonumber \\
&\quad
= {}- \frac{2}{\omega} \int d\beta\, d\alpha\,
    \frac{\alpha f^g(\beta,\alpha)}{\alpha\xi + \beta} \,,
\end{align}
which is zero because $\alpha f^g(\beta,\alpha)$ is odd in both
$\beta$ and in $\alpha$.  The $D$-term contribution to $H^g$ gives
\begin{equation}
\mathcal{I}^g(\omega) = \frac{2}{\omega} \int_{-1}^1 d\alpha\,
  \frac{D^g(\alpha)}{\omega-\alpha}
\end{equation}
in agreement with \eqref{D-C} and \eqref{I-g-C}.  For $F^g = E^g$ one
finds analogous results with the opposite sign for $C^g$ and $D^g$.
One thus obtains integral relations
\begin{align}
  \label{int-g-relations}
& \int_{-1}^1 \frac{dx}{x}\,
  \biggl[ H^{g}(x,\xi)
        - H^{g}\Bigl(x,\frac{x}{\omega}\Bigr) \biggr]
  \left[ \frac{1}{\omega\xi - x} - \frac{1}{\omega\xi + x}
  \right]
\nonumber \\
&\quad = - \int_{-1}^1 \frac{dx}{x}\,
  \biggl[ E^{g}(x,\xi)
        - E^{g}\Bigl(x,\frac{x}{\omega}\Bigr) \biggr]
  \left[ \frac{1}{\omega\xi - x} - \frac{1}{\omega\xi + x}
  \right]
= \frac{2}{\omega} \int_{-1}^1 dx\, \frac{D^g(x)}{\omega-x}
\end{align}
and dispersion representations
\begin{align}
  \label{disp-rel-5g}
\re\mathcal{H}^{g}(\xi) &=
  \frac{1}{\pi} \int_1^\infty d\omega\, \im C^{g}(\omega)
  \int_{-1}^1 dx\, \biggl\{ \frac{1}{x}\ms
    H^{g}\Bigl(x,\frac{x}{\omega}\Bigr)\,
    \left[ \frac{1}{\omega\xi - x} - \frac{1}{\omega\xi + x}
    \right]
  + \frac{2}{\omega}\, \frac{D^g(x)}{\omega-x} \biggr\} \,,
\nonumber \\
\re\mathcal{E}^{g}(\xi) &=
  \frac{1}{\pi} \int_1^\infty d\omega\, \im C^{g}(\omega)
  \int_{-1}^1 dx\, \biggl\{ \frac{1}{x}\ms
    E^{g}\Bigl(x,\frac{x}{\omega}\Bigr)\,
    \left[ \frac{1}{\omega\xi - x} - \frac{1}{\omega\xi + x}
    \right]
  - \frac{2}{\omega}\, \frac{D^g(x)}{\omega-x} \biggr\} \,.
\end{align}
Furthermore one finds
\begin{equation}
\lim_{\xi\to\infty} \mathcal{H}^g(\xi)
= - \lim_{\xi\to\infty} \mathcal{E}^g(\xi)
= \frac{1}{\pi} \int_1^\infty \frac{d\omega}{\omega}\,
  \im C^{g}(\omega)\, \int_{-1}^1 dx\; \frac{2 D^g(x)}{\omega-x}
\end{equation}
for the invariant amplitudes at $\xi\to \infty$.  We remark in passing
that \eqref{int-g-relations} and \eqref{disp-rel-5g} may be rewritten
using
\begin{equation}
\frac{1}{x}
  \left[ \frac{1}{\omega\xi - x} - \frac{1}{\omega\xi + x} \right]
= \frac{1}{\omega \xi}
  \left[ \frac{1}{\omega\xi - x} + \frac{1}{\omega\xi + x} \right] \,.
\end{equation}


\subsection{Polarized gluon distributions}
\label{sec:pol-g}

Let us now discuss the generalized gluon distributions in the
polarized sector, which appear in DVCS starting at NLO in $\alpha_s$.
As in the previous section we begin with the factorization formula
\eqref{fact-g}, where now $F^g(x,\xi) = \widetilde{H}^g(x,\xi)$,
$\widetilde{E}^g(x,\xi)$ is odd in $x$.  The hard-scattering kernel
$C^g$ is even in $\omega$ and vanishes like $\omega^2$ for $\omega\to
0$.  The invariant amplitudes $\widetilde{\mathcal{H}}^{g}(\xi)$ and
$\widetilde{\mathcal{E}}^{g}(\xi)$ have negative signature.  The NLO
calculation of $C^{g}(\omega)$ for DVCS gives a large-$\omega$
behavior like $\omega^{-1}$ up to logarithms, and higher orders will
have the same power behavior as discussed in the first paragraph of
Sect.~\ref{sec:pol-q}.

Assuming a small-$x$ behavior $x \Delta g(x) \to 0$ of the polarized
gluon density, which is required for the existence of the moment
$\int_0^1 dx\, \Delta g(x)$ and consistent with global fits of parton
densities, we expect that $\smash{\xi \widetilde{\mathcal{H}}^{g}(\xi)
\to 0}$ for $\xi\to 0$.  We then readily obtain dispersion relations
as in \eqref{disp-rel-3} and \eqref{disp-rel-4} with the replacements
$\mathcal{F}^{q[-]} \to \mathcal{\widetilde{H}}^g$, $C^{q[-]} \to C^g$
and $F^{q[-]} \to 2 x^{-1} \widetilde{H}^g$.  Their consistency
requires the $\xi$-independence of
\begin{equation}
  \label{I-gtilde-def}
\mathcal{I}^{g}(\omega)
= \int_{-1}^1 \frac{dx}{x}\,
  \biggl[ \widetilde{H}^{g}(x,\xi)
        - \widetilde{H}^{g}\Bigl(x,\frac{x}{\omega}\Bigr) \biggr]
  \left[ \frac{1}{\omega\xi - x} + \frac{1}{\omega\xi + x}
  \right] \,,
\end{equation}
where the principal value prescription is to be taken at $x=0$.
As in Sect.~\ref{sec:unpol-g} we can rewrite this as
\begin{align}
  \label{not-yet-zero}
\mathcal{I}^{g}(\omega)
&= \frac{2}{\omega} \sum_{n=1}^\infty \frac{1}{n!}\,
   \Bigl( \frac{\partial}{\partial\eta} \Bigr)^n
   \int_{-1}^1 \frac{dx}{x}\,
     \Bigl(\frac{x}{\omega}-\xi\Bigr)^{n-1}
     \widetilde{H}^g(x,\eta)\, \Bigr|_{\eta=\xi}
\nonumber \\
&= \frac{2}{\omega} \sum_{n=1}^\infty \frac{1}{n!}\, (-\xi)^{n-1}\,
   \Bigl( \frac{\partial}{\partial\eta} \Bigr)^n
   \int_{-1}^1 \frac{dx}{x}\,
     \widetilde{H}^g(x,\eta)\, \Bigr|_{\eta=\xi} \,,
\end{align}
where in the second step we have expanded the factor $(x/\omega
-\xi)^{n-1}$ and used the polynomiality properties \eqref{gpd-g-mom}
of $\widetilde{H}^g$.  To proceed we need to know the dependence of
$\int dx\, x^{-1} \widetilde{H}^g(x,\eta)$ on $\eta$.

In \cite{Diehl:2003ny} a double distribution representation for
$\widetilde{H}^g$ was given, which has the same form as \eqref{dd-g2}
for $H^g_f$.  Inserting this into \eqref{I-gtilde-def} one obtains an
expression as in \eqref{I-g-dd}, which is nonzero because the
corresponding double distribution is even and not odd in $\beta$.
Such a double distribution representation for $\widetilde{H}^g$ (as
well as its analog for $\widetilde{E}^g$) is however incomplete,
because for the $x^{n-2}$ moment of the distributions it gives a
polynomial with highest power $\xi^{n-3}$ (with $n$ being odd) instead
of $\xi^{n-1}$ as required in \eqref{gpd-g-mom}.  To obtain a correct
representation, we can use the construction discussed in
\cite{Belitsky:2000vk} for the generalized quark distribution in the
pion.  This leads to writing a double distribution representation for
$x^{-1} \widetilde{H}^g$ and $x^{-1} \widetilde{E}^g$, i.e.
\begin{align}
  \label{dd-gtilde}
\widetilde{H}^g(x,\xi,t) &= x \int d\beta\, d\alpha\;
  \delta(x-\alpha\xi-\beta)\, \widetilde{f}^g(\beta,\alpha,t) \,,
\nonumber \\
\widetilde{E}^g(x,\xi,t) &= x \int d\beta\, d\alpha\;
  \delta(x-\alpha\xi-\beta)\, \widetilde{k}^g(\beta,\alpha,t) \,,
\end{align}
where $\widetilde{f}^g$ and $\widetilde{k}^g$ are even in $\alpha$ and
$\beta$.  We note that in the forward limit $t=0$ one has $\int
d\alpha\, \widetilde{f}^g(x,\alpha,0) = \Delta g(x)$, which is much
less singular than the corresponding limit $x^{-1}\ms q_\pi(x)$ for
the double distribution of quarks in the pion considered in
\cite{Belitsky:2000vk} and should thus be less problematic for the
purpose of model building.

Apart from giving the required maximum power of $\xi^{n-1}$ for the
$x^{n-2}$ moments of $\widetilde{H}^g$ and $\widetilde{E}^g$, the
representation \eqref{dd-gtilde} also has the important consequence
that
\begin{equation}
  \label{neg-mom}
\int_{-1}^1 \frac{dx}{x}\, \widetilde{H}^g(x,\eta) =
  \int d\beta\, d\alpha\; \widetilde{f}^g(\beta,\alpha)
\end{equation}
is independent of $\eta$, so that according to 
\eqref{not-yet-zero} 
\begin{equation}
  \mathcal{I}^g(\omega) = 0
\end{equation}
is independent of $\xi$, which we had to show.  This is also seen by
direct insertion of \eqref{dd-gtilde} into \eqref{I-gtilde-def}, which
leads to an expression of the form \eqref{int-H-dd} we encountered for
quark distributions.  We thus finally obtain dispersion
representations as in \eqref{disp-rel-tilde} with the replacements
$\widetilde{\mathcal{H}}^{q[-]} \to \widetilde{\mathcal{H}}^g$,
$C^{q[-]} \to C^g$ and $\widetilde{H}^{q} \to x^{-1} \widetilde{H}^g$,
as well as the limit $\widetilde{\mathcal{H}}^g(\xi) \to 0$ for
$\xi\to\infty$.

For the invariant amplitude $\widetilde{\mathcal{E}}^g$ we must take
into account a possible spin-zero exchange in the $t$-channel
(although the exchange of an $\eta$ or $\eta'$ in the flavor singlet
sector is most likely not of the same phenomenological importance as
pion exchange in $\widetilde{E}^q$).  With the double distribution
representation \eqref{dd-gtilde} one can proceed exactly as for the
case of quark distributions in Sect.~\ref{sec:pol-q}.  One thus
obtains analogs of the dispersion representations
\eqref{disp-rel-tilde} with the replacements
$\widetilde{\mathcal{H}}^{q[-]} \to \widetilde{\mathcal{E}}^g$,
$C^{q[-]} \to C^g$ and $\widetilde{H}^{q} \to x^{-1} \widetilde{E}^g$,
as well as the results
\begin{equation}
\int_{-1}^1 \frac{dx}{x}\, \biggl[ \widetilde{E}^{g}(x,\xi)
        - \widetilde{E}^{g}\Bigl(x,\frac{x}{\omega}\Bigr) \biggr]
  \left[ \frac{1}{\omega\xi - x} + \frac{1}{\omega\xi + x}
  \right] = 0
\end{equation}
and
\begin{align}
  \label{Etilde-asy-g}
\lim_{\xi\to\infty} \xi \widetilde{\mathcal{E}}^{g}(\xi) &=
  \frac{1}{\pi} \int_1^\infty d\omega\, \im C^{g}(\omega)\,
  \sum_{\substack{n=1\\{\mathrm{odd}}}}^\infty
  \left(\frac{2}{\omega}\right)^{n} 2 \widetilde{B}^g_{n,n-1} \,.
\end{align}
To avoid a $\delta(x)$ contribution in $x^{-1} \widetilde{E}^g(x,
x/\omega)$ due to spin-zero exchange one may use the analog of
\eqref{et-alt}, which reads
\begin{align}
  \label{et-alt-g}
\re \xi \widetilde{\mathcal{E}}^{g}(\xi)
= \lim_{\xi_0 \to\infty} \xi_0\ms \widetilde{\mathcal{E}}^{g}(\xi_0)
 + \frac{1}{\pi} \int_1^\infty \frac{d\omega}{\omega}\,
   \im C^{g}(\omega) \int_{-1}^1 dx\,
     \widetilde{E}^{g}\Bigl(x,\frac{x}{\omega}\Bigr)\,
     \left[ \frac{1}{\omega\xi - x}
            - \frac{1}{\omega\xi + x} \right] \,.
\end{align}


\subsection{Helicity-flip distributions}
\label{sec:flip}

We conclude this section with a few remarks on the generalized parton
distributions for quark or gluon helicity flip, which have been
introduced and discussed in \cite{Hoodbhoy:1998vm,Diehl:2001pm}.

In the quark case these distributions are chiral-odd, and to date
there is no simple exclusive process known where they appear.
Reactions like $\gamma^* p \to \rho\ms \rho\ms p$ were proposed in
\cite{Ivanov:2002jj}, but due to their three-particle final state the
discussion of dispersion relations would be much more complicated.
However, integral relations analogous to \eqref{rel-tilde} are valid
for the quark distributions $H_T^q$, $E_T^q$, $\widetilde{H}_T^q$ and
$\widetilde{E}_T^q$ defined in \cite{Diehl:2001pm}.  As we saw in
Sect.~\ref{sec:gpd-conseq}, their derivation only requires the
$x^{n-1}$ moments of the distributions to be polynomials in $\xi$ with
maximal power $\xi^{n-1}$.  This is indeed the case, as has been shown
in \cite{Hagler:2004yt}.

Gluon helicity-flip distributions appear in DVCS starting at order
$\alpha_s$, with the hard-scattering formula of the form
\begin{equation}
  \label{fact-gT}
\mathcal{F}_T^{g}(\xi)
= \int_{-1}^1 dx\, \frac{1}{\xi}\,
  C_T^{g}\Bigl( \frac{x}{\xi} \Bigr)\,
  \frac{F_T^{g}(x,\xi)}{x}
\end{equation}
for $F_T^g = H^g_T, E^g_T, \widetilde{H}^g_T, \widetilde{E}^g_T$ as
defined in \cite{Diehl:2001pm}.  Dispersion representations for this
case can be discussed in analogy to the cases considered in the
previous sections.  To do this requires analysis of the high-energy
behavior (see the related work \cite{Ermolaev:1998jv} for the
helicity-flip structure function $F_3^\gamma$ of the photon) and of
the polynomiality properties (in generalization of the quark case
treated in \cite{Hagler:2004yt}).  We shall not do this here.


\section{The model of Freund, McDermott and Strikman}
\label{sec:mc}

As an application of the dispersion relations discussed in this work,
we now investigate the model for GPDs proposed by Freund, McDermott
and Strikman in \cite{Freund:2002qf}.  We focus on the quark singlet
distribution and its generalized counterpart,
\begin{align}
\Sigma(x) &= \sum_q \bigl[\ms q(x) + \bar{q}(x) \ms\bigr] \,,
&
H(x,\xi) &= \sum_q H^{q[+]}(x,\xi) \,,
\end{align}
where for ease of notation we have not explicity indicated that
$H(x,\xi)$ refers to the quark singlet.  Here and in the following we
take $t=0$, which does not affect the issue of analyticity to be
discussed.  In our notation, the model introduced in
\cite{Freund:2002qf} reads
\begin{equation}
  \label{mc-model-q}
H(x,\xi) =
\begin{cases}
\Sigma(x)  & \text{for $x\geq \xi$} \\[0.3em]
{\displaystyle \Sigma(\xi)\; \frac{x}{\xi}
\left[ 1 + \frac{15}{2}\ms a(\xi) \left(1-\frac{x^2}{\xi^2}\right)
\right]}     & \text{for $x< \xi$}
\end{cases}
\end{equation}
with $a(\xi)$ chosen to satisfy the polynomiality condition
\begin{equation}
\int^1_0 dx\, x\ms H(x,\xi)
 = \sum_q \int_{-1}^1 dx \, x\ms H^q(x,\xi)
 = \int^1_0 dx \, x \ms \Sigma(x) + 4\xi^2 C_2
\end{equation}
for the lowest nontrivial Mellin moment, where $C_2 = \sum_q
C^q_2(t=0)$ according to \eqref{gpd-q-mom}.  One readily finds
\begin{equation}
  \label{mc-aq}
\Sigma(\xi)\, a(\xi) = \frac{1}{\xi^2} \int^\xi_0 dx\, x\ms \Sigma(x)
  - \frac{1}{3}\ms \Sigma(\xi) + 4\ms C_2 \,.
\end{equation}
Clearly, higher Mellin moments of \eqref{mc-model-q} are generally not
polynomials in $\xi$ of the order required by \eqref{gpd-q-mom}.  At
small $\xi$, one may expect that this does not have an important
effect on the moments themselves, in the sense that a Taylor expansion
\begin{equation}
\int_0^1 dx\, x^{n-1}\, H(x,\xi)
=\sum_{\substack{k=0\\{\mathrm{even}}}}^{\infty}
 (2\xi)^k\, A_{n,k}
\end{equation}
of a given moment differs from a polynomial of order $\xi^n$ by terms
vanishing like $\xi^{n+2}$ for $\xi\to 0$.  It is however not obvious
that this only leads to small inconsistencies in scattering amplitudes
calculated with \eqref{mc-model-q}, given that these do not have a
simple expression in terms of Mellin moments with integer index $n$.

We have seen that polynomiality of the Mellin moment ensures the
consistency of dispersion relations for the hard-scattering kernel and
for the process amplitude.  Let us check by how much the dispersion
representations \eqref{disp-rel-3} and \eqref{disp-rel-4} differ for
the above model.  We limit ourselves to the lowest order in $\alpha_s$
and take $\im C^{q[+]}(\omega) = \pi\ms \bigl[\ms \delta(\omega-1) -
\delta(\omega+1) \ms\bigr]$, omitting any global factors in the
kernel.  The two dispersion representations then read
\begin{align}
  \label{disp-test}
\re \mathcal{H}_{dir}(\xi) =
  \int_0^1 dx\, & H(x,\xi)
  \left[ \frac{1}{\xi - x} - \frac{1}{\xi + x} \right] \,,
\nonumber \\
\re \mathcal{H}_{\xi_0}(\xi) =
  \int_0^1 dx\, & \biggl\{ H(x,x)\,
  \left[ \frac{1}{\xi - x} - \frac{1}{\xi + x} \right]
+ \biggl[ H(x,\xi_0) - H(x,x) \biggr]
    \left[ \frac{1}{\xi_0 - x} - \frac{1}{\xi_0 + x} \right]
\biggr\} \,.
\end{align}
We note that at Born level $\re \mathcal{H}_{dir}(\xi)$ calculated
from \eqref{disp-rel-3} coincides with the real part calculated
directly from the factorization formula \eqref{fact-form}.
For a numerical study, we take
\begin{equation}
  \label{q-param}
x \Sigma(x) =
  p_1\ms  x^{-p_2} (1-x)^{\ms p_3} (1 + p_4\ms x)
\end{equation}
for the quark singlet distribution, with $p_1=0.34$, $p_2=0.25$,
$p_3=4$, $p_4=25.4$.  This gives a reasonably good approximation of
the CTEQ6M distributions at scale $\mu=2 \gev$.  With $p_3$ taken as
an integer, the integrals required for evaluating \eqref{mc-aq} and
\eqref{disp-test} are readily carried out.  One finds that $\re
\mathcal{H}_{dir}(\xi_0)$ diverges for $\xi_0 \to\infty$ in this
model, so that one cannot use this point for the subtraction required
in $\re \mathcal{H}_{\xi_0}$.  We take instead the $s$-channel
threshold $\xi_0 = 1$, where the model GPD has the simple form $H(x,1)
\propto x \ms (1-x^2)$.  As an alternative choice we take the value
$\xi_0 = 0.01$ in the small-$\xi$ region.  The comparison of the two
representations in \eqref{disp-test} for several values of $\xi$ is
given in Table~\ref{tab:mc}.  We see that their discrepancy is severe
and does not improve with decreasing $\xi$.  By construction, the two
representations coincide of course for $\xi=\xi_0$.

\begin{table}
  \caption{\label{tab:mc} The convolution integrals $\re
    \mathcal{H}_{dir}(\xi)$ and $\re \mathcal{H}_{\xi_0}(\xi)$
    defined in \protect\eqref{disp-test}, evaluated for $\xi_0=1$ and
    $\xi_0=0.01$.  The values are calculated with the GPD model
    specified by \protect\eqref{mc-model-q} and \protect\eqref{mc-aq}
    with $C_2=0$.  For better legibility, the values of the integrals
    have been rounded to two significant digits in the first two rows
    and to the next integer in the remaining ones.}
\renewcommand{\arraystretch}{1.4}
\begin{center}
\begin{tabular}{cccccc} \hline\hline
$\xi$ & $\re \mathcal{H}_{dir}$
      & $\re \mathcal{H}_{1.0}$
      & $\re \mathcal{H}_{\, 0.01} \rule[-1.3em]{0pt}{3.2em}$ 
& $\displaystyle\frac{\re \mathcal{H}_{1.0}}{\re \mathcal{H}_{dir}}$
& $\displaystyle\frac{\re \mathcal{H}_{\, 0.01}}{\re \mathcal{H}_{dir}}$
\\ \hline
$10^{-4}$ & $~12 \times 10^{4}$ & $4.4 \times 10^{4}$ & $4.4 \times 10^{4}$
          & $0.37$ & $0.37$  \\
$10^{-3}$ & $6.5 \times 10^{3}$ & $2.3 \times 10^{3}$ & $2.5 \times 10^{3}$
          & $0.35$ & $0.39$ \\
$10^{-2}$ & $318$ & $74$ & $318$  & $0.23$ & $1$ \\
$0.1$ & $26$ & $9$  & $253$ & $0.37$ & $10$ \\
$0.3$ & $16$ & $11$ & $255$ & $0.70$ & $16$ \\
$0.5$ & $10$ & $7$  & $251$ & $0.76$ & $26$ \\
\hline\hline
\end{tabular}
\end{center}
\end{table}

The values in the table have been obtained by setting $C_2$ to zero in
\eqref{mc-aq}.  One readily finds that this term gives a contribution
of $20\ms C_2$ to both $\re \mathcal{H}_{dir}(\xi)$ and $\re
\mathcal{H}_{\xi_0}(\xi)$.  Taking the value of $C_2 \approx -0.8$
estimated in the chiral quark-soliton model \cite{Goeke:2001tz} would
not significantly change the values for small $\xi$, and in any case
cannot restore the discrepancy between the two integrals in
\eqref{disp-test}.  We must conclude that, even for small $\xi$, the
model \eqref{mc-model-q} violates polynomiality and thus Lorentz
invariance in a way which leads to serious inconsistencies when using
it to calculate the real part of process amplitudes.  To obtain
consistent results, one may use the ansatz \eqref{mc-model-q} for $|x|
\ge \xi$ to calculate $\im \mathcal{H}(\xi)$ and to restore the real
part from the dispersion relation \eqref{disp-rel-2}, with the
subtraction constant left undetermined by the model.


\section{Summary}
\label{sec:sum}

Lorentz invariance implies that the Mellin moments of generalized
parton distributions are polynomials in the skewness $\xi$ with a
maximal power depending on the quantum numbers of the distribution.  We
have shown that this property leads to integral relations
\begin{align}
  \label{int-rel-fin}
\int_{-1}^1 dx\, F(x,\xi,t) 
  \left[ \frac{1}{\omega\xi - x} -\sigma \frac{1}{\omega\xi + x} \right]
= \int_{-1}^1 dx\, F\Bigl( x,\frac{x}{\omega},t \Bigr)
  \left[ \frac{1}{\omega\xi - x} -\sigma \frac{1}{\omega\xi + x} \right]
+ \mathcal{I}(\omega,t)
\end{align}
for $\sigma=\pm 1$ and any $\omega \ge 1$, where $F$ is one of the
distributions
\begin{equation}
H^q, E^q, \widetilde{H}^q, \widetilde{E}^q,
\quad
\frac{H^g}{x}, \frac{E^g}{x}, 
  \frac{\widetilde{H}^g}{x}, \frac{\widetilde{E}^g}{x}, 
\quad
H_T^q, E_T^q, \widetilde{H}_T^q, \widetilde{E}_T^q.
\end{equation}
In \eqref{int-rel-fin} Cauchy's principal value prescription is to be
used at $x=\pm \omega\xi$ and at $x=0$.
The only cases where $\mathcal{I}(\omega,t)$ is nonzero occur for
unpolarized distributions and $\sigma=+1$, where
\begin{align}
  \label{D-term-fin}
{}\pm \mathcal{I}(\omega,t) &=
  2 \sum_{\substack{n=2\\{\mathrm{even}}}}^\infty
    \left(\frac{2}{\omega}\right)^{n} C^q_{n}(t)
= 2 \int_{-1}^1 dx\, \frac{D^q(x,t)}{\omega-x}
& & \text{for~} F = H^q, E^q,
\displaybreak
\nonumber \\
{}\pm \mathcal{I}(\omega,t) &=
  4 \sum_{\substack{n=2\\{\mathrm{even}}}}^\infty
    \left(\frac{2}{\omega}\right)^{n} C^g_{n}(t)
= \frac{2}{\omega} \int_{-1}^1 dx\, \frac{D^g(x,t)}{\omega-x}
& & \text{for~} F = \frac{H^g}{x}, \frac{E^g}{x}.
\end{align}
Here the sign $+$ on the l.h.s.\ is to be taken for $H^q$, $H^g$ and
the sign $-$ for $E^q$, $E^g$.  To establish the relations
\eqref{int-rel-fin} in the polarized gluon sector, we needed that the
moments $\int dx\, x^{-1} \widetilde{H}^g(x,\xi,t)$ and $\int dx\,
x^{-1} \widetilde{E}^g(x,\xi,t)$ are independent of $\xi$, and we had
to correct the double distribution representation of $\widetilde{H}^g$
and $\widetilde{E}^g$ used so far in the literature.

For $t\le 0$ the real part of the leading invariant amplitudes for
DVCS or meson production can be obtained from a dispersion relation of
the hard-scattering kernel,
\begin{align}
\re\mathcal{F}(\xi,t) &= \frac{1}{\pi} \int_1^{\infty} d\omega\,
  \im C(\omega)\, \int_{-1}^1 dx\, F(x,\xi,t) 
  \left[ \frac{1}{\omega\xi - x} -\sigma \frac{1}{\omega\xi + x}
  \right] \,,
\intertext{or for the invariant amplitude itself,}
\re\mathcal{F}(\xi,t) &= \frac{1}{\pi} \int_1^{\infty} d\omega\,
  \im C(\omega)\, \biggl\{\,
  \int_{-1}^1 dx\, F\Bigl( x,\frac{x}{\omega},t \Bigr)
  \left[ \frac{1}{\omega\xi - x} -\sigma \frac{1}{\omega\xi + x} \right]
  + \mathcal{I}(\omega,t) \biggr\} \,,
  \label{disp-rel-fin}
\end{align}
where $C = C^{q [\sigma]}, C^g$ is the appropriate hard-scattering
kernel (for the quark transversity distributions no corresponding
process is known).  Consistency of the two representations is ensured
by \eqref{int-rel-fin}.
The contribution from $\mathcal{I}(\omega,t)$ in \eqref{disp-rel-fin}
is energy independent and can be identified with $\mathcal{F}(\xi,t)$
in the limit $\xi\to \infty$, i.e.\ at the point $2\nu=s-u=0$ below
threshold.  The corresponding terms given in \eqref{D-term-fin} are
due to spin-zero exchange in the $t$-channel.
Spin-zero exchange contributions in the parity-odd sector appear in
$\widetilde{E}^q$ and $\widetilde{E}^g$.  They do not give a nonzero
$\mathcal{I}(\omega,t)$ but can instead generate a term proportional
to $\delta(x)$ in $F(x, x/\omega, t)$.  In the alternative dispersion
representations \eqref{et-alt} and \eqref{et-alt-g} for $\re
\xi\mathcal{E}^{q[-]}(\xi,t)$ and $\re \xi\mathcal{E}^g(\xi,t)$ such a
$\delta(x)$ term is avoided, and the spin-zero exchange contribution
appears directly as a subtraction constant, with
\begin{equation}
  \label{spin-zero-fin}
\sum_{\substack{n=1\\{\mathrm{odd}}}}^\infty
  \left(\frac{2}{\omega}\right)^{n} \widetilde{B}^q_{n,n-1}(t)
\qquad \text{or} \qquad
2\, \sum_{\substack{n=1\\{\mathrm{odd}}}}^\infty
  \left(\frac{2}{\omega}\right)^{n} \widetilde{B}^g_{n,n-1}(t)
\end{equation}
playing the same role as $\mathcal{I}(\omega,t)$ in
\eqref{disp-rel-fin}.

In Sect.~\ref{sec:mc} we have seen that the relation
\eqref{int-rel-fin} can be strongly violated in models of GPDs that do
not respect polynomiality, even for small $\xi$.  In particular, we
found that the model proposed in \cite{Freund:2002qf} leads to serious
conflicts with dispersion relations when used for calculating the real
part of scattering amplitudes.

The representation \eqref{disp-rel-fin} has important consequences on
the information about GPDs that can be extracted from DVCS and meson
production.  To leading approximation in $\alpha_s$, the imaginary
part of the amplitude is only sensitive to the distributions at
$x=\xi$, and the only additional information contained in the real
part is a constant associated with pure spin-zero exchange, given by
\eqref{D-term-fin} or \eqref{spin-zero-fin} at $\omega=1$.  In
\cite{Teryaev:2005uj} this was referred to as a holographic property.
Beyond leading order the evaluation of both imaginary and real parts
of the amplitude involves however the full DGLAP region $|x| \ge \xi$.
In addition, the real part depends on the appropriate spin-zero term
in \eqref{D-term-fin} or \eqref{spin-zero-fin} at all $\omega \ge 1$.
We remark that in \cite{Kumericki:2007sa} the possibility was
discussed to reconstruct the subtraction terms in \eqref{D-term-fin}
from the imaginary part of the DVCS amplitude combined with the
inclusive deep inelastic cross section.

Consider the comparison of a given model or parameterization of GPDs
with data on DVCS or meson production.  In a leading-order analysis
(which should of course always be restricted to kinematics where the
LO approximation is adequate) it is sufficient to characterize each
GPD by its values at $x=\xi$, supplemented by a constant for the
spin-zero exchange contribution discussed above.  On one hand this can
be a welcome simplification, and on the other hand it indicates the
limitations of an LO analysis: when confronting data with a given GPD
one is sensitive to $x \neq\xi$ (and to the details of the spin-zero
exchange contribution) only at NLO or higher accuracy.

Let us finally emphasize that the imaginary part of an amplitude
involves GPDs with skewness given by the value of $\xi$ in the
measurement, whereas the dispersion representation
\eqref{disp-rel-fin} of the real part involves all values of the
skewness from 0 to 1.  For measurements in a limited energy region,
the extra information of the real part compared with the imaginary one
is thus not limited to the spin-zero exchange terms.


\section*{Acknowledgments}

We gratefully acknowledge discussions with I. Anikin, D. M\"uller and
A. Sch\"afer.  D.~I.\ thanks the DESY Theory Group and the Institute
for Theoretical Physics at the University of Regensburg for their
hospitality.  This work is supported by the Helmholtz Association,
contract number VH-NG-004, and the work of D.~I.\ is supported in part
by grants RFBR-06-02-16064 and NSh 5362.2006.2.


\end{document}